\documentclass[%
 aps,
 pre,
 amsmath,amssymb,
 reprint
]{revtex4-1}

\usepackage[utf8]{inputenc}  % UTF-8 input
\usepackage[T1]{fontenc}     % proper font encoding
\usepackage[english]{babel}  % document language

\usepackage{mathptmx}        % Times font for text + math

\usepackage{graphicx}        % include figures
\usepackage{dcolumn}         % align table columns on decimal
\usepackage{bm}              % bold math symbols
\usepackage[usenames,dvipsnames]{color} % colored text

\usepackage{hyperref}        % clickable references
\hypersetup{
    colorlinks=true,
    linkcolor=blue,
    citecolor=blue,
    urlcolor=blue
}

\usepackage{listings}        % code listings (if needed)
\usepackage{etoolbox}
\makeatletter
\def\@email#1#2{%
 \endgroup
 \patchcmd{\titleblock@produce}
  {\frontmatter@RRAPformat}
  {\frontmatter@RRAPformat{\produce@RRAP{*#1\href{mailto:#2}{#2}}}\frontmatter@RRAPformat}
  {}{}
}%
\makeatother

\begin{document}

\title{Large Differences Between Stochastic and Deterministic Kinetics in a Simple Autocatalytic Reaction Network}

\author{Tomasz Bednarek}
\author{Jakub Jędrak}
\email{jjedrak@ichf.edu.pl}
\affiliation{Institute of Physical Chemistry, Polish Academy of Sciences, Kasprzaka 44/52, 01-224 Warsaw, Poland}

\date{\today}

\begin{abstract}

In small systems, quantitative discrepancies between stochastic and deterministic descriptions of chemical kinetics can be significant, with their magnitude depending on the specific reaction network. Here, we study the Finke–Watzky model—an irreversible autocatalysis, $\mathrm{A} + \mathrm{B} \rightarrow 2\mathrm{B}$, supplemented by an irreversible first-order process, $\mathrm{A} \rightarrow \mathrm{B}$. This model has been used to describe the formation of transition metal nanoparticles and protein misfolding and aggregation, but it may also serve as a minimal model for the spread of a non-fatal but incurable disease. We show that, for certain parameter values, exceptionally large deviations can arise between stochastic and deterministic kinetics of the Finke–Watzky model. Moreover, its stochastic time evolution may be highly sensitive to initial conditions. These properties are retained in the generalization of the model to reversible reactions. To quantify the differences between the predictions of deterministic and stochastic kinetics, we derive the explicit analytical solution of the Chemical Master Equation for the Finke–Watzky model. This solution also allows us to derive analogous solutions for two related reaction networks: $\mathrm{A} + \mathrm{A} \rightarrow \mathrm{A} + \mathrm{B}$, $\mathrm{A} \rightarrow \mathrm{B}$, and $\mathrm{A} + \mathrm{A} \rightarrow \mathrm{A} + \mathrm{B}$, $\mathrm{A} + \mathrm{B} \rightarrow 2\mathrm{B}$. Our findings may have implications for modeling epidemics and intracellular chemical processes, and more broadly for models of population dynamics.

\end{abstract}

\maketitle
\vspace{-8ex}

\section{Introduction} %C

The kinetics of chemical reactions in spatially homogeneous systems can be described either by deterministic rate equations \cite{houston2012chemical, smith2013basic} or by stochastic approaches, the most important of which is the Chemical Master Equation (CME) \cite{mcquarrie1967stochastic}. Alternatively, Gillespie’s Stochastic Simulation Algorithm (SSA) \cite{gillespie1976general, gillespie1977exact, gillespie1992rigorous, gillespie2007stochastic} yields results that are, in principle, equivalent to those of the CME. Approaches based on the CME or SSA are more fundamental and provide a more realistic description than deterministic kinetics, which neglect the discrete nature of molecules (or individuals) and the effectively stochastic character of intermolecular collisions. These considerations, as well as the following discussion, apply more generally to population dynamics, for example to models of epidemic spreading.

Differences between the predictions of stochastic and deterministic models can be substantial, both quantitatively and qualitatively \cite{togashi2001transitions}. For example, the number and stability of steady states may differ: some states may be stable in the deterministic framework but unstable in the stochastic one or vice versa \cite{elgart2004rare, khasin2012minimizing, ma2012small, biancalani2012noise, biancalani2014noise, smith2016extinction}.

In small systems, where the discreteness of molecule numbers becomes significant, quantitative differences are also expected: the time evolution of the average number of molecules predicted by the CME or SSA may deviate markedly from the deterministic concentration profiles \cite{mcquarrie1967stochastic, gillespie1977exact}. Only in the case of purely first-order reactions does the mean molecule number follow the same time dependence as the corresponding deterministic concentration \cite{gillespie2007stochastic}. For more complex reactions, deterministic kinetics are recovered only when fluctuations in molecule numbers are neglected.

As system size increases, the discrepancies between deterministic and stochastic predictions are expected to vanish \cite{kurtz1972relationship}. However, the rate at which this convergence occurs—i.e., how large the system must be to justify a deterministic description—depends on the specific reaction network. For example, it has been shown \cite{schuster2019special} that in the simple autocatalytic reaction $\mathrm{A} + \mathrm{B} \rightarrow 2\mathrm{B}$, the discrepancies between the average number of molecules and the deterministic concentrations can be much larger than in the bimolecular reaction $\mathrm{A} + \mathrm{B} \rightarrow 2\mathrm{C}$, even when the total number of molecules is the same and of order one hundred. In Ref. \cite{schuster2019special}, this effect was termed "stochastic delay", since, in the case of autocatalysis, stochastic descriptions such as the CME and SSA yield noticeably slower reaction dynamics than deterministic kinetic models.

Here, we show that this stochastic delay becomes even more pronounced when the autocatalytic reaction is supplemented by an irreversible first-order process, $\mathrm{A} \rightarrow \mathrm{B}$,
\begin{align}
\label{1st_reaction_WF}
\mathrm{A} &\xrightarrow{k_1}  \mathrm{B}, \\
\label{2nd_reaction_WF}
\mathrm{A} + \mathrm{B} &\xrightarrow{k_2}  2\mathrm{B}.
\end{align}
The stochastic delay and the fluctuations around the mean are especially significant when the system is initially free of $\mathrm{B}$ and when $k_1 \ll k_2$. For this reason, unlike in Ref.~\cite{schuster2019special}, we do not use SSA to compute the stochastic dynamics of the system—Gillespie’s standard algorithm becomes ineffective when reaction rates differ greatly, and advanced variants such as tau-leaping \cite{gillespie2001approximate, cao2005avoiding, cao2006efficient, cao2007adaptive} are not fully equivalent to the CME. Therefore, we derive an explicit analytical solution of the CME for the reactions (\ref{1st_reaction_WF})–(\ref{2nd_reaction_WF}). We are not aware of any prior publication of such a result.

Such a solution is significant in its own right, as the reaction set (\ref{1st_reaction_WF})–(\ref{2nd_reaction_WF}), known as the Finke–Watzky model (FWM) \cite{watzky1997transition}, is used to describe various important phenomena. Originally introduced as a minimalistic “Ockham’s razor” model of transition metal colloid formation \cite{watzky1997transition}, it remains widely used for that purpose (see, e.g., Refs.~\cite{yang2017autocatalytic, luty2024synthesis}). In this context, $\mathrm{A}$ is a soluble metal complex (e.g., $[\mathrm{Au}\mathrm{Cl}_4]^-$), and $\mathrm{B}$ represents all metal nanoparticles, irrespective of size or shape. Typically, both reactions (\ref{1st_reaction_WF}) and (\ref{2nd_reaction_WF}) are pseudoelementary, consisting of multiple elementary steps.\footnote{For instance, reaction (\ref{1st_reaction_WF}) is actually bimolecular, involving a reducing agent. However, with a large excess of the reducing agent, its concentration remains effectively constant, rendering the reaction pseudo-first-order.} The FWM has also been extended to model nanoparticles of various sizes \cite{jkedrak2013exact, jedrak2014cluster}.

The reaction network (\ref{1st_reaction_WF})–(\ref{2nd_reaction_WF}) also serves as a minimal model of protein misfolding and aggregation \cite{watzky2008fitting, morris2008fitting, morris2009protein, iashchishyn2017finke}, processes implicated in prion and neurodegenerative diseases such as Alzheimer’s and Parkinson’s. Here, $\mathrm{A}$ is a native protein, while $\mathrm{B}$ denotes misfolded, catalytically active aggregates.

Lastly, the FWM can be interpreted as a simple model for the spread of an incurable but non-fatal disease. Healthy individuals ($\mathrm{A}$) may become infected either through contact with infected individuals ($\mathrm{B}$) or indirectly (via air, surfaces, vectors, etc.).

For all of the above phenomena, the first-order process (\ref{1st_reaction_WF}) is indispensable, and therefore autocatalysis (\ref{2nd_reaction_WF}) alone is not sufficient as a minimal effective model.\footnote{An assessment of how realistic the Finke–Watzky model is for describing the phenomena and processes discussed here, and to what extent it can serve as their minimal, effective quantitative description, lies beyond the scope of the present work; see, however, Ref. \cite{jkedrak2013exact,jedrak2014cluster}.}

The deterministic rate equations for reactions (\ref{1st_reaction_WF})–(\ref{2nd_reaction_WF}) are straightforward to solve and have been used to fit experimental data on nanoparticle formation \cite{watzky1997transition} as well as on protein misfolding and aggregation \textit{in vitro} \cite{watzky2008fitting, morris2008fitting, morris2009protein, iashchishyn2017finke}. However, some applications of the FWM may involve relatively small systems—for example, modeling protein misfolding and aggregation \textit{in vivo} or disease spread in small populations—where stochastic approaches such as the CME or SSA are more appropriate. The existence of significant differences between stochastic and deterministic descriptions of this model further motivates deriving an analytical solution of the CME.

While our focus is on irreversible reactions, we also analyze the reversible generalization of the FWM:
\begin{eqnarray}
\label{1st_reaction_WF_rev}
\mathrm{A} & \underset{k^{-}_{1}}{\stackrel{k^{+}_{1}}{\rightleftharpoons}} & \mathrm{B}, \\
\label{2nd_reaction_WF_rev}
\mathrm{A} + \mathrm{B} & \underset{k^{-}_{2}}{\stackrel{k^{+}_{2}}{\rightleftharpoons}} & \mathrm{B} + \mathrm{B}.
\end{eqnarray}
The reaction network (\ref{1st_reaction_WF_rev})–(\ref{2nd_reaction_WF_rev}) with $k^{-}_{2} = 0$ has been proposed as a modification of the FWM for modeling silver nanoparticle formation with sodium borohydride as the reducing agent \cite{amirjani2018modified}. It is also applicable to modeling diseases with possible recovery ($\mathrm{B} \rightarrow \mathrm{A}$). In the most general case ($k^{-}_{2} > 0$), this set of chemical equations also corresponds to a Lindemann-type mechanism, in which $\mathrm{B}$ is the initial reactant molecule and $\mathrm{A}$ is an activated reaction intermediate.

Here, we study (\ref{1st_reaction_WF_rev})–(\ref{2nd_reaction_WF_rev}) primarily to examine how the presence of the inverse reactions affects the stochastic delay. However, for this reversible network, an analytical time-dependent solution of the CME is not available; numerical methods must be used instead. Numerical results show, as expected, that for a range of values of the rate constants for the inverse reactions ($k^{-}_{1}$ and $k^{-}_{2}$) large differences between stochastic and deterministic solutions still persist. For the reaction set (\ref{1st_reaction_WF_rev})–(\ref{2nd_reaction_WF_rev}), we also derive analytical formulas for the steady-state probability distribution.

Finally, the analytical solution of the CME for the FWM also allows us to obtain, by a simple mapping of the model parameters, analogous solutions for two related reaction networks: $\mathrm{A} + \mathrm{A} \rightarrow \mathrm{A} + \mathrm{B}$, $\mathrm{A} \rightarrow \mathrm{B}$ (which we refer to here as the 'inverse Finke–Watzky model') and $\mathrm{A} + \mathrm{A} \rightarrow \mathrm{A} + \mathrm{B}$, $\mathrm{A} + \mathrm{B} \rightarrow 2\mathrm{B}$ (see Section~\ref{Two_other_reactions_Appendix}). This is notable because analytical solutions of the CME are rare \cite{mcquarrie1967stochastic}. Exact solutions are known so far for the dimerization reaction $2 \mathrm{A} \rightarrow \mathrm{C}$ \cite{mcquarrie1967stochastic, laurenzi2000analytical}, for the simple bimolecular reaction $\mathrm{A} + \mathrm{B} \rightarrow 2 \mathrm{C}$ \cite{laurenzi2000analytical}, for pure autocatalysis $\mathrm{A} + \mathrm{B} \rightarrow 2\mathrm{B}$ \cite{arslan2008kinetics}, and for a few simple one-step irreversible birth–death processes \cite{lee2012analytical}. To our knowledge, explicit analytical solutions of the CME for networks involving two or more reactions, at least one of which is bimolecular, have not been reported. This work fills also that gap.

\section{Theoretical Framework \label{Theory}} 

We assume the system under consideration to be spatially homogeneous and closed, with no exchange of molecules with the environment (e.g., well‑stirred reagents in a batch reactor). Consequently, the total molecule number is conserved:
\begin{equation}
\label{constraint_on_mass_conservation}
N_{A}(t) + N_{B}(t) \equiv M = \mathrm{const}, \quad t \ge 0,
\end{equation}
and the mesoscopic state is fully determined by either $N_{A}(t)$ or $N_{B}(t)$. Without loss of generality, we track only $N_A(t)$.

\subsection{Irreversible reactions \label{irreversible_reactions_theory}}

Both the Finke–Watzky model (\ref{1st_reaction_WF})–(\ref{2nd_reaction_WF}) and the two other sets of irreversible reactions considered here are one‑step processes. Their Chemical Master Equation (CME) takes the general form \cite{van2007stochastic}:
\begin{equation}
\label{CME_WF}
\frac{dP_n(t)}{dt} = r_{n+1}\,P_{n+1}(t) \;-\; r_{n}\,P_{n}(t), 
\quad n = 0,1,\dots,N,
\end{equation}
where $P_{n}(t)$ is the probability of finding $n$ molecules of A (and $M-n$ of B) at time $t$, $N = N_A(0)$, and $M-N = N_B(0)$. We set $P_n(t)=0$ for $n>N$ and $r_0=0$. The rate coefficients $r_n$ depend on the particular reaction network. We assume the deterministic initial condition
\begin{equation}
\label{deterministic_initial_condition}
P_{n}(0)=\delta_{nN}.
\end{equation}
More general initial conditions can be included without difficulty, if required. 

Equation~(\ref{CME_WF}) can be written in matrix form \cite{van2007stochastic},
\begin{equation}
\label{CME_WF_matrix_form}
\frac{d\mathbf{P}(t)}{dt} = \mathbb{W}\,\mathbf{P}(t),
\quad
\mathbf{P}(t) = [P_0(t),P_1(t),\ldots,P_N(t)]^T,
\end{equation}
with
\begin{equation}
\label{W_matrix_explicit_form}
\mathbb{W} = 
\begin{pmatrix}
0      & r_1    & 0      & \cdots & 0      \\
0      & -r_1   & r_2    & \cdots & 0      \\
\vdots & \ddots & \ddots & \ddots & \vdots \\
0      & \cdots & 0      & -r_{N-1} & r_N  \\
0      & \cdots & 0      & 0       & -r_N 
\end{pmatrix}.
\end{equation}
Since $\mathbb{W}$ is upper triangular, its eigenvalues are $\lambda_n=-r_n$.

Because we are dealing with irreversible one-step reactions, there are no terms involving $P_{n-1}(t)$ in Eq.~(\ref{CME_WF}), and one can obtain an analytical solution via the Laplace transform (Appendix \ref{Solution_of_CME_WF_Appendix}). Assuming nondegenerate eigenvalues, one finds
\begin{align}
\label{CME_WF_solution_N_t_main}
P_N(t) &= e^{-r_N t}, \\
\label{CME_WF_solution_n_t_main}
P_n(t) &= \sum_{k=n}^N 
\frac{\prod_{j=n+1}^N r_j}{\prod_{\substack{j=n\\j\neq k}}^N (r_j - r_k)}
\,e^{-r_k t}
\;\equiv\;\sum_{k=n}^N C_{nk}\,e^{-r_k t},
\end{align}
for $n=1,\dots,N-1$. $P_0(t)$ follows from normalization or directly from Eq.~(\ref{CME_WF}) at $n=0$. Here $C_{NN}=1$. Finally, since $r_0=0$ and $r_n>0$ for $n\ge1$, the steady‐state solution is
\begin{equation}
\label{deterministic_final_condition}
P_n^{(s)} = \lim_{t\to\infty}P_n(t) = \delta_{n0}.
\end{equation}

The solution (\ref{CME_WF_solution_N_t_main})--(\ref{CME_WF_solution_n_t_main}) of Eqs. (\ref{CME_WF}) has appeared in various contexts and has been derived independently by different authors \cite{allnatt1968theory, arslan2008kinetics, lee2012analytical, rudnicki2014modele, jedrak2014cluster}. In contrast, by "explicit analytical solution" we mean here not the general form (\ref{CME_WF_solution_N_t_main})--(\ref{CME_WF_solution_n_t_main}) of the solution, but the most concise analytical form of both $P_n(t)$ and its lowest moments that can be obtained for a specific choice of the coefficients $r_n$.

%\section{Stationary solution of the CME for reversible reactions}\mathbf{P}(t)

\subsection{Reversible reactions}

In the case of reversible processes (\ref{1st_reaction_WF_rev})-(\ref{2nd_reaction_WF_rev}), instead of Eq. (\ref{CME_WF}) we have the CME of a more general form \cite{van2007stochastic}
\begin{equation}
\label{CME_WF_rev}
\frac{dP_n(t)}{dt} = r_{n+1}P_{n+1}(t) - (r_{n} + g_{n})P_{n}(t) + g_{n-1}P_{n-1}(t), 
\end{equation}
where $n \in \{0, \dots M\}$, $ M \equiv N_{A}(0) + N_{B}(0)$ defined in Eq. (\ref{constraint_on_mass_conservation}) is a total number of molecules in the system, and $P_{n}(t)=0$ for $n \geqslant M+1 $. $r_n$ and $g_n$ are the kinetic rate coefficients, with $r_0 = g_{-1} = g_M = r_{M+1} = 0$. Equation (\ref{CME_WF_rev}) can be rewritten as
\begin{eqnarray}
\label{CME_WF_matrix_form_rev}
\frac{d \mathbf{P}(t)}{dt} = \mathbb{W}^{\prime} \mathbf{P}(t),
\end{eqnarray}
where now
\begin{equation}
\label{W_matrix_explicit_form_rev}
\mathbb{W}^{\prime} = 
\begin{pmatrix}
-g_0 & r_1  & 0 &  0 &\cdots & 0 \\
g_0 & -(r_1 + g_1) & r_2 &   &\cdots  & 0\\
0 & g_1    & -(r_2 + g_2) & r_3  &\cdots & 0 \\
0 & 0    & g_2 & -(r_3 + g_3)  &\cdots &  0 \\
\vdots  &  \vdots & \vdots  &\vdots  & \ddots  & \vdots  \\
0 & 0 & 0 & 0 &\cdots   & r_M \\ 
0 & 0 &  0& 0 &\cdots & -r_M 
\end{pmatrix}.
\end{equation}
%
%$P_n(t)$ 
% 
For reversible bimolecular reactions in general, and for (\ref{1st_reaction_WF_rev})–(\ref{2nd_reaction_WF_rev}) in particular, the functional form of $\mathbf{P}(t)$ can no longer be determined analytically, and we must resort to numerical methods. This may involve either SSA simulations or a direct, exact numerical solution of the system of ODEs (\ref{CME_WF_matrix_form_rev}) or a method of Ref.~\cite{smith2015general}. 

However, the equation (\ref{CME_WF_rev}) describes the time evolution of a one-step process, so its stationary solution ($dP_n(t)/dt = 0$) is easy to find \cite{van2007stochastic, lee2012analytical}. We have 
\begin{equation}
\label{CME_WF_rev_stationary_solution_det_bal}
r_{n} P^{(s)}_n  = g_{n-1} P^{(s)}_{n-1} 
\end{equation}
and therefore 
\begin{equation}
\label{CME_WF_rev_stationary_solution}
P^{(s)}_n = P^{(s)}_0 \prod_{j=0}^{n-1}\frac{g_j}{r_{j+1}}. 
\end{equation}
$P^{(s)}_0$ can be determined from the normalization condition \cite{van2007stochastic}:
\begin{eqnarray}
\label{CME_WF_rev_stationary_solution_normalization}
\frac{1}{P^{(s)}_0} &=& 1 + \sum_{n=1}^M  \prod_{j=0}^{n-1}\frac{g_j}{r_{j+1}}.
\end{eqnarray}

\section{Results \label{Results}}

In this section, we present our results. First, we derive the explicit analytical solution of the CME for the original formulation of the Finke--Watzky model (FWM) (\ref{1st_reaction_WF})--(\ref{2nd_reaction_WF}). We then use this solution to analyze the differences between stochastic and deterministic kinetics for selected choices of model parameters and initial conditions (Subsection \ref{when_are_SD}). The approximate approach based on a truncated set of moment equations is discussed in Subsection \ref{method_of_moments}.

Next, in Subsection \ref{Results_rev} we consider a generalization of the FWM to reversible reactions (\ref{1st_reaction_WF_rev})--(\ref{2nd_reaction_WF_rev}). In Subsection \ref{Results_inv}, we present the explicit time-dependent solution of the CME for the 'inverse Finke--Watzky model' (\ref{1st_reaction_inv_WF})--(\ref{2nd_reaction_inv_WF}), whereas in Subsection \ref{Results_two_consecutive} we discuss the analogous solution for two consecutive bimolecular reactions (\ref{1st_reaction_consecutive})--(\ref{2nd_reaction_consecutive}).

\subsection{The Finke-Watkzy model $\mathrm{A} \rightarrow \mathrm{B}$, $\mathrm{A} + \mathrm{B} \rightarrow 2\mathrm{B}$ \label{Results_irr}}

For the reaction set (\ref{1st_reaction_WF})-(\ref{2nd_reaction_WF}), the coefficients $r_{n}$ appearing in (\ref{CME_WF}), (\ref{W_matrix_explicit_form}), (\ref{CME_WF_solution_N_t_main}) and (\ref{CME_WF_solution_n_t_main}) are given by 
\begin{eqnarray}
\label{eigenvalues_of_CME_WF}
r_{n} = - \lambda_n =  n [k_1 +  k_2(M-n)] =  k_2 n(\tilde{M} - n),
\end{eqnarray}
where we define the following quantities
\begin{eqnarray}
\label{parameters_of_CME_WF_definitions}
\tilde{M} \equiv  M+\nu,  ~~~~~~ \nu \equiv \frac{k_1}{k_2},
\end{eqnarray}
and where $ M \equiv N_{A}(0) + N_{B}(0)$ has been defined in Eq. (\ref{constraint_on_mass_conservation}). The spectrum $\{0, \lambda_1, \ldots, \lambda_{N-1}, \lambda_N \}$ (\ref{eigenvalues_of_CME_WF})  of the matrix $\mathbb{W}$ (\ref{W_matrix_explicit_form}) is not degenerate if $\tilde{M} \notin \mathbb{N}$, i.e., if  $\nu$ is not an integer. If $\tilde{M} \in \mathbb{N}$, some eigenvalues may be doubly degenerate.% depending on the values of . 

In particular, when $k_1 = 0$, the FWM (\ref{1st_reaction_WF})–(\ref{2nd_reaction_WF}) reduces to a simple autocatalytic reaction, for which the degeneracy of the eigenvalues (\ref{eigenvalues_of_CME_WF}) of $\mathbb{W}$  has been analyzed in Ref. \cite{arslan2008kinetics}. However, in the present context it is legitimate to disregard the degeneracy of the spectrum (\ref{eigenvalues_of_CME_WF}), since the values of $\nu$ that lead to degenerate $\lambda_n$ form a set of measure zero. Moreover, any $\nu \in \mathbb{N} \cup {0}$ can be treated as the limit of a sequence of non-integer values of $\nu$. Therefore, the presence of an additional first-order reaction (\ref{1st_reaction_WF}) effectively removes the degeneracy, and thus makes solving the CME for the FWM easier than in the case of pure autocatalysis (\ref{2nd_reaction_WF}).\footnote{In Ref. \onlinecite{arslan2008kinetics} the non-physical parameter $\epsilon$, which lifts the degeneracy, was introduced for practical convenience. The final results were obtained by taking the limit $\epsilon \to 0$.}

\subsubsection{Probability distribution and its two lowest moments}

After some algebra, we find that for the $r_{n}$ given by (\ref{eigenvalues_of_CME_WF}) with non-integer $\tilde{M}$, the explicit form of the coefficient $C_{nk}$ defined by (\ref{CME_WF_solution_n_t_main}) is
\begin{widetext}
\begin{eqnarray}
\label{explicit_form_of_the_C_n_k_coefficients}
C_{nk} = (-1)^{N+k} \frac{(2k - \tilde{M}) \Gamma(N+1) \Gamma(\tilde{M}- n) \Gamma(n + k - \tilde{M})}{\Gamma(\tilde{M}-N)\Gamma(N - k + 1) \Gamma(N + k + 1 - \tilde{M}) \Gamma(k-n +1) \Gamma(n+1)}.
\end{eqnarray}
Together with (\ref{CME_WF_solution_N_t_main}), (\ref{CME_WF_solution_n_t_main}), and (\ref{eigenvalues_of_CME_WF}), equation (\ref{explicit_form_of_the_C_n_k_coefficients}) determines the time dependence of the probability distribution:
\begin{eqnarray}
\label{CME_WF_solution_n_t_main_explicite}
P_{n}(t) & = & \sum_{l=n}^{N} (-1)^{N+l} {\binom{N}{l}}{\binom{l}{n}} \frac{ (2l - \tilde{M}) \Gamma(\tilde{M}- n) \Gamma(n + l - \tilde{M})}{\Gamma(\tilde{M}-N) \Gamma(N + l + 1 - \tilde{M}) } \exp\left[- k_2 l (\tilde{M} - l) t\right].
\end{eqnarray}
\end{widetext}
Using (\ref{CME_WF_solution_n_t_main_explicite}), we can now obtain the time evolution of the two lowest raw moments of $P_n(t)$,
\begin{eqnarray}
\label{first_moment_non_deg_WF}
\mu_{1A}(t) &\equiv & \sum_{n=0}^N n P_{n}(t) = \sum_{l=1}^{N} E^{(1)}_l e^{-k_2 l (\tilde{M} - l) t}, \nonumber \\
E^{(1)}_l  & = &  \frac{N! }{(N-l)!} \frac{ (2l - \tilde{M})\Gamma(N + 1 - \tilde{M})}{\Gamma(N + l + 1 - \tilde{M})} \nonumber \\ & = &  (-1)^l \frac{N!}{(N-l)!} \frac{ (2l - \tilde{M}) \Gamma(\tilde{M} - N - l)}{\Gamma(\tilde{M} - N)}
\end{eqnarray}
and
\begin{eqnarray}
\label{second_moment_non_deg_WF}
\mu_{2A}(t) &\equiv & \sum_{n=0}^N n^2 P_{n}(t) = \sum_{l=1}^{N} E^{(2)}_l e^{-k_2 l (\tilde{M} - l) t}, \nonumber \\
E^{(2)}_l & = &  (l^2 - \tilde{M}l + {\tilde{M}}) E^{(1)}_l.
\end{eqnarray}
The details of the derivation of (\ref{first_moment_non_deg_WF}) and (\ref{second_moment_non_deg_WF}) are given in Appendix~\ref{Derivation_of_moments_app}. Higher moments can be obtained in a similar way.

The average number of B molecules is given by $\mu_{1B}(t) = \sum_{n=0}^N (M-n) P_{n}(t) = M - \mu_{1A}(t)$, while the second moments are related by $\mu_{2B}(t) = \mu_{2A}(t) - 2 M \mu_{1A}(t) + M^2$. Therefore, the standard deviation $\sigma(t)$ of the corresponding probability distribution is the same for both A and B molecules,
\begin{equation}
\label{sigma_definition}
\sigma(t) \equiv \sqrt{\mu_{2A}(t) - \mu_{1A}^2(t)} = \sqrt{\mu_{2B}(t) - \mu_{1B}^2(t)}.
\end{equation}
Analytical expressions for the moments (\ref{first_moment_non_deg_WF})–(\ref{second_moment_non_deg_WF}) reduce the double sum (which arises when calculating the moment directly from its definition using $P_n(t)$ given by Eq.~\ref{CME_WF_solution_n_t_main_explicite}) to a single sum, lowering the computational cost. A caveat is that numerical evaluation of Eqs.~(\ref{first_moment_non_deg_WF}) and (\ref{second_moment_non_deg_WF}) requires arbitrary-precision arithmetic: for $M=100$ the absolute values of the coefficients $E^{(1)}_l$ and $E^{(2)}_l$ reach $10^{30}$ and $10^{33}$, respectively, so small relative round-off errors at short times can cause large absolute errors in the moments.\footnote{Here, the GNU Multiple Precision Floating-Point Reliable Library (GNU MPFR) was used for this purpose (\url{https://www.mpfr.org}).} In fact, instead of relying on the potentially cumbersome arbitrary-precision evaluation of Eqs.~(\ref{first_moment_non_deg_WF})–(\ref{second_moment_non_deg_WF}), one can simply solve the set of $N+1$ ODEs in Eq.~(\ref{CME_WF}) by direct numerical integration. We performed the calculations using both approaches and obtained perfect agreement.

\subsubsection{When is the stochastic delay most pronounced in the Finke–Watzky model? \label{when_are_SD}}

As with other reaction networks, for the FWM (\ref{1st_reaction_WF})–(\ref{2nd_reaction_WF}) we expect that deviations between the deterministic and stochastic descriptions become more pronounced as the total number $M$ of molecules decreases. For sufficiently small $M$, different realizations of the reaction starting from the same initial condition can diverge markedly (large fluctuations around the average number of molecules), as reflected in a high standard deviation $\sigma(t)$ (\ref{sigma_definition}).

However, the reaction network under consideration is exceptional in the following sense. Even for relatively large $M$ (e.g., $M=100$), significant systematic deviations can arise between the deterministic concentration $a(t)$ (see Eq.~\ref{Det_Chem_Kin_WF_only_a_solution_of} in Appendix~\ref{relation_of_sto_to_det}) and its stochastic counterpart,  
\begin{equation} 
\label{stochastic_concentration} 
a_s(t)=\frac{\mu_{1A}(t)}{V},
\end{equation}
depending on the initial conditions and rate constants. Here $\mu_{1A}(t)$ (\ref{first_moment_non_deg_WF}) is the average number of A molecules, and $V$ is the total volume of the system.

\begin{figure}[tb]
\begin{center}					  				
\rotatebox{270}{\scalebox{0.34}{\includegraphics{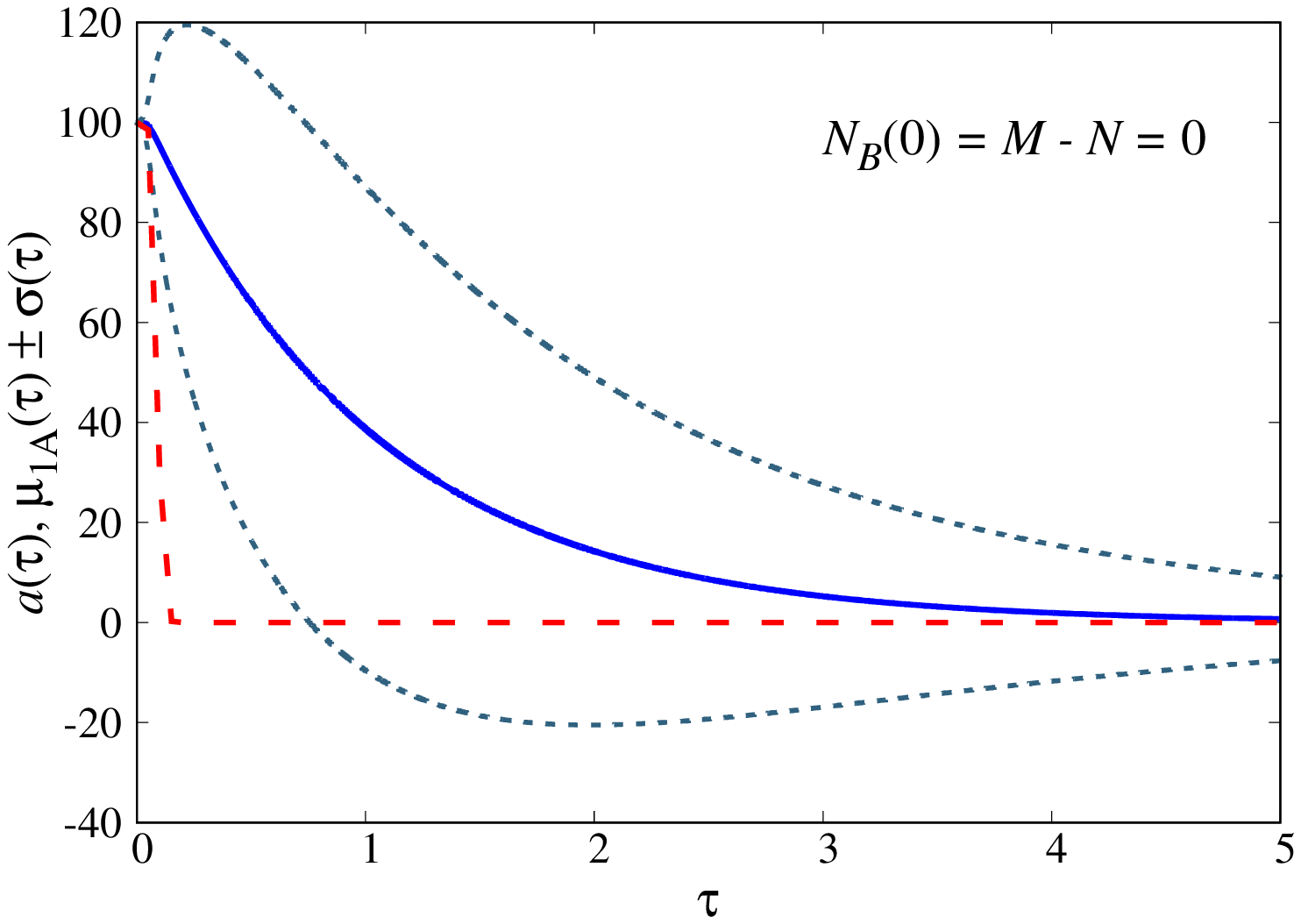}}} 
\end{center}  
\caption{Stochastic versus deterministic kinetics for the reactions (\ref{1st_reaction_WF})--(\ref{2nd_reaction_WF}), with initial condition $N_B(0) = 0$, and rate constants $k_1 = 0.001$, $k_2 = 0.1$. The concentration of A given by  $a(t)$ (\ref{Det_Chem_Kin_WF_only_a_solution_of}) is shown as a red dashed line, and decays significantly faster than its stochastic counterpart $a_s(t) = \mu_{1A}(t)/V$ (\ref{stochastic_concentration}), plotted as a thick blue solid line. The light blue dotted lines indicate the single standard deviation envelope around the mean, $\mu_{1A}(t) \pm \sigma(t)$, providing a quantitative measure of stochastic fluctuations. The system volume is set to $V = 1$, so that $k_2 = \mathcal{K}_2$ and $a_s(t) = \mu_{1A}(t)$, with $\mu_{1A}(t)$ given by Eq.~(\ref{first_moment_non_deg_WF}). All quantities are plotted as functions of the dimensionless time variable $\tau = k_2 t$.}
\label{sd_WF_nu_0p01_M_100_N_100_up_label}
\end{figure}

\begin{figure}[h]
\begin{center}					  				
\rotatebox{270}{\scalebox{0.34}{\includegraphics{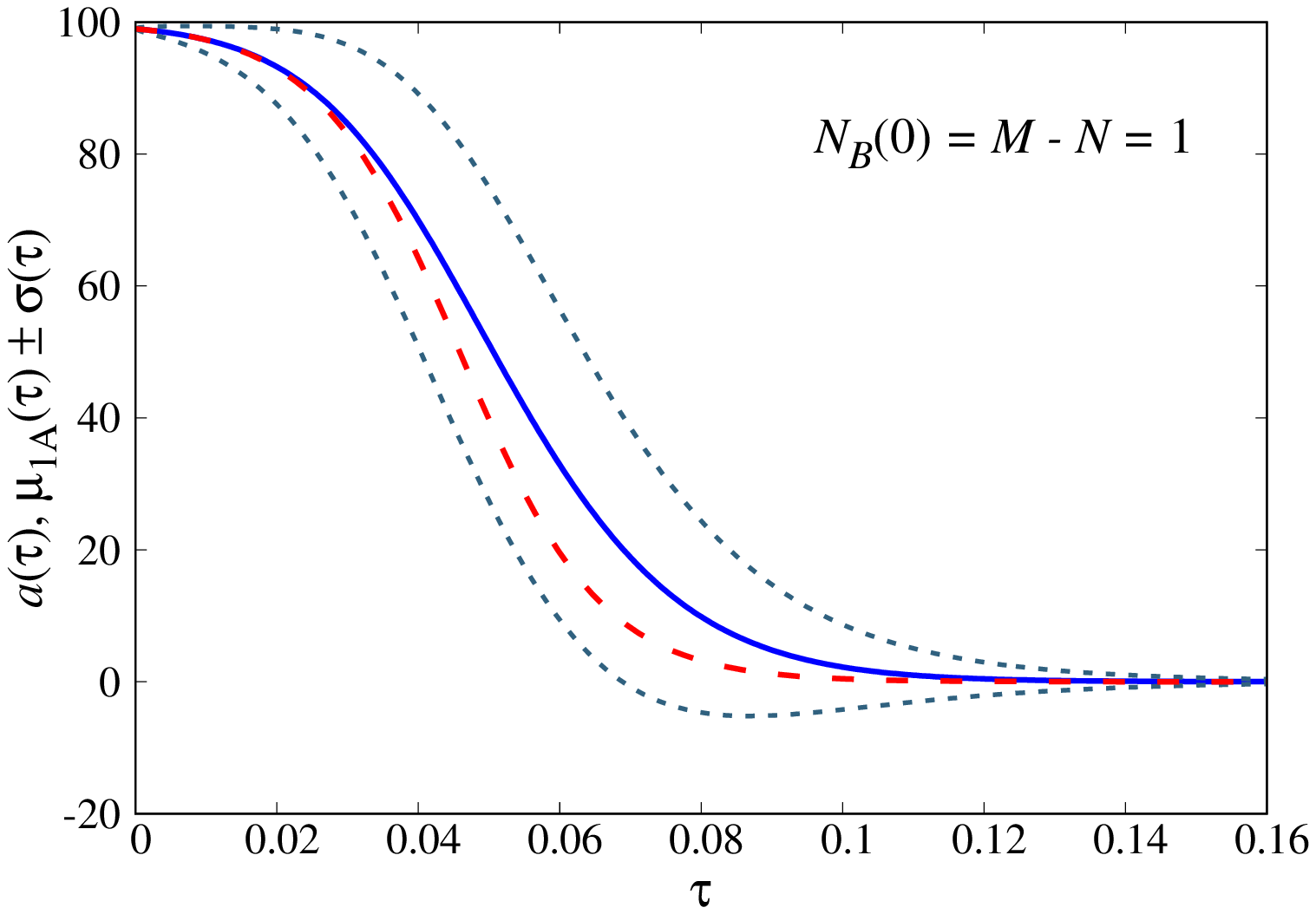}}} 
\end{center}  
\caption{Stochastic versus deterministic kinetics for reactions (\ref{1st_reaction_WF})–(\ref{2nd_reaction_WF}) with $N_B(0)=1$, $k_1 = 0.001$, and $k_2 = 0.1$. When a single B molecule is initially present, the difference between the deterministic concentration $a(t)$ (\ref{Det_Chem_Kin_WF_only_a_solution_of}) (red dashed line) and its stochastic counterpart $a_s(t) = \mu_{1A}(t)/V$ (\ref{stochastic_concentration}) (thick blue solid line) is much smaller than in Fig.~\ref{sd_WF_nu_0p01_M_100_N_100_up_label}. Fluctuations around the mean, $\mu_{1A}(t) \pm \sigma(t)$ (light blue dotted lines), are also significantly reduced compared to that figure. Here, $V=1$ so that $k_2 = \mathcal{K}_{2}$ and $a_s(t) = \mu_{1A}(t)$, with $\mu_{1A}(t)$ from (\ref{first_moment_non_deg_WF}). All quantities are plotted as functions of the dimensionless time $\tau = k_2 t$. Note the different $\tau$ range compared to Fig.~\ref{sd_WF_nu_0p01_M_100_N_100_up_label}.}
\label{sd_WF_nu_0p01_M_100_N_99_up_label}
\end{figure}

These deviations become particularly pronounced when the ``nucleation'' step [Eq.~(\ref{1st_reaction_WF})] is much slower (and thus much less likely) than the autocatalytic reaction [Eq.~(\ref{2nd_reaction_WF})], and when no $B$ molecules are present initially., i.e.,
\begin{equation}
\label{sto_det_max_conditions}
M - N = N_B(0) = 0 
\quad\text{and}\quad 
k_1 \ll k_2.
\end{equation}
This scenario is illustrated in Fig.~\ref{sd_WF_nu_0p01_M_100_N_100_up_label}, where we compare $a(t)$ [Eq.~\ref{Det_Chem_Kin_WF_only_a_solution_of}] with the analytical expression for $\mu_{1A}(t)$ [Eq.~\ref{first_moment_non_deg_WF}] (here we set $V=1$). If instead we begin with a single B molecule ($N_B(0)=1$) while keeping $k_1$ and $k_2$ fixed, the discrepancy between $\mu_{1A}(t)$ and $a(t)$ is much reduced (Fig.~\ref{sd_WF_nu_0p01_M_100_N_99_up_label}).

%%---------------END OF FIGURE------------------%
%
%%---------------END OF FIGURE------------------%
%

%
\begin{figure}[h]
\begin{center}					  				
\rotatebox{270}{\scalebox{0.34}{\includegraphics{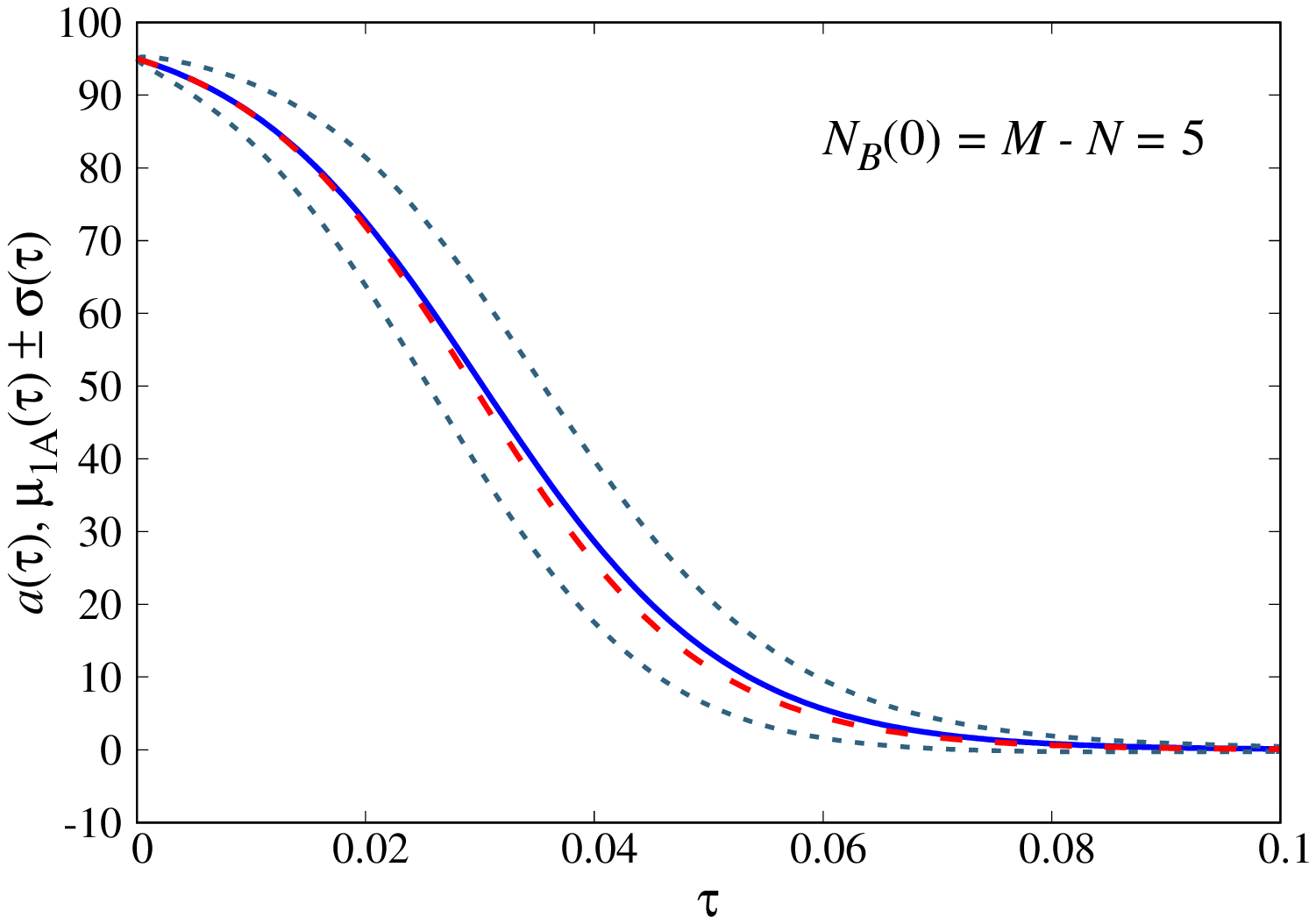}}} 
\end{center}  
\caption{Stochastic versus deterministic kinetics for reactions (\ref{1st_reaction_WF})–(\ref{2nd_reaction_WF}) with $N_B(0)=5$, $k_1 = 0.001$, and $k_2 = 0.1$. The difference between the deterministic concentration $a(t)$ (\ref{Det_Chem_Kin_WF_only_a_solution_of}) (red dashed line) and its stochastic counterpart $a_s(t) = \mu_{1A}(t)/V$ (\ref{stochastic_concentration}) (thick blue solid line) is smaller than in Fig.~\ref{sd_WF_nu_0p01_M_100_N_99_up_label}. All quantities are plotted as functions of the dimensionless time $\tau = k_2 t$.}
\label{sd_WF_nu_0p01_M_100_N_95_up_label}
\end{figure}
%%---------------END OF FIGURE------------------%
%
%
\begin{figure}[h]
\begin{center}					  				
\rotatebox{270}{\scalebox{0.34}{\includegraphics{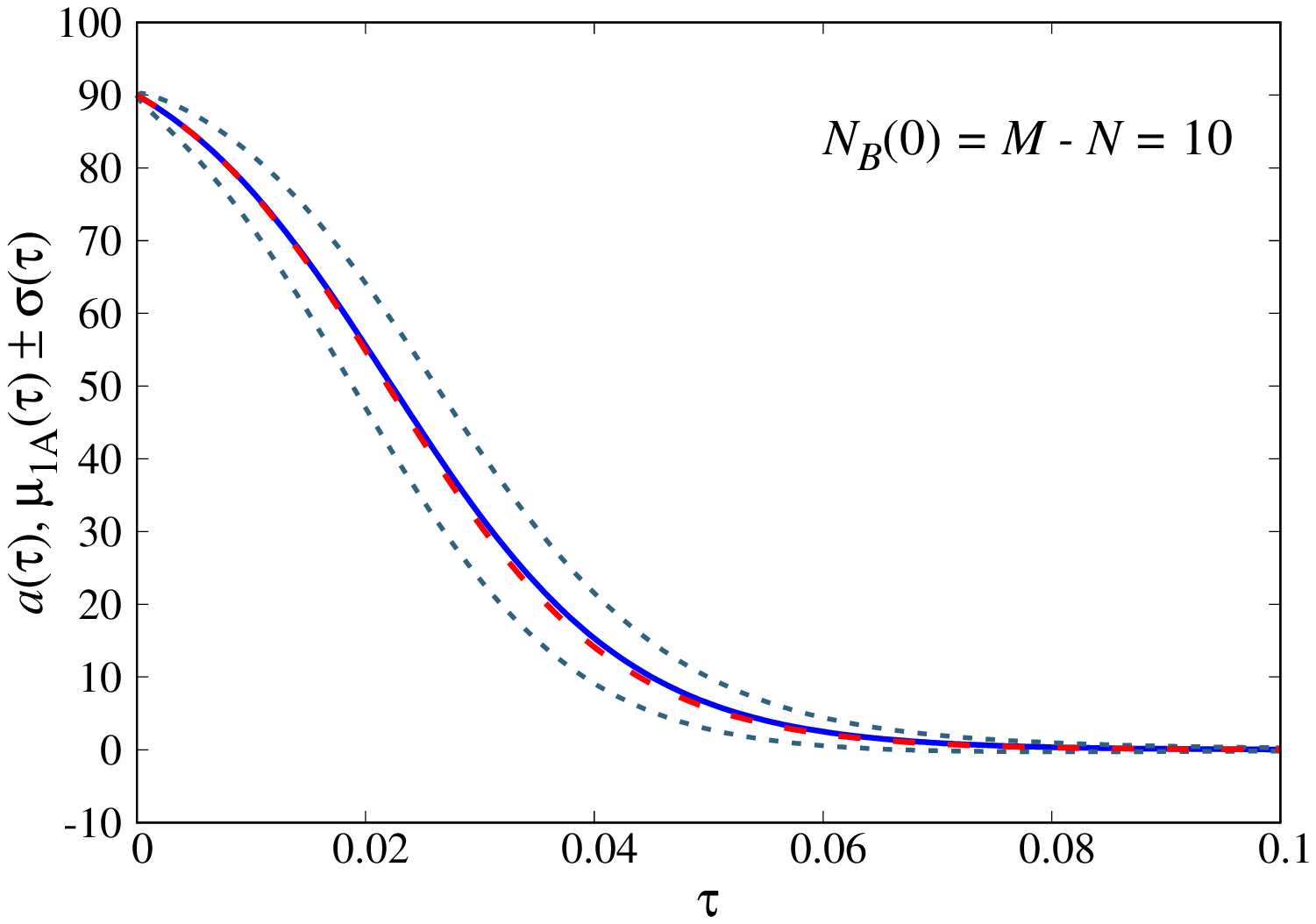}}} 
\end{center}  
\caption{Stochastic versus deterministic kinetics for reactions (\ref{1st_reaction_WF})–(\ref{2nd_reaction_WF}) with $N_B(0)=10$, $k_1 = 0.001$, and $k_2 = 0.1$. The difference between the deterministic concentration $a(t)$ (\ref{Det_Chem_Kin_WF_only_a_solution_of}) (red dashed line) and its stochastic counterpart $a_s(t) = \mu_{1A}(t)/V$ (\ref{stochastic_concentration}) (thick blue solid line) is now negligible. Fluctuations around the mean (light blue dotted lines) are comparable to those in Fig.~\ref{sd_WF_nu_0p01_M_100_N_90_up_label}. All quantities are plotted as functions of the dimensionless time $\tau = k_2 t$.}
\label{sd_WF_nu_0p01_M_100_N_90_up_label}
\end{figure}

As shown in Figures \ref{sd_WF_nu_0p01_M_100_N_95_up_label} and \ref{sd_WF_nu_0p01_M_100_N_90_up_label}, the discrepancies between stochastic and deterministic descriptions diminish as the initial number of B molecules, $N_B(0) = M - N$, increases. Even at $N_B(0) = 5$, these differences are quite small (Fig.~\ref{sd_WF_nu_0p01_M_100_N_95_up_label}), and become practically negligible by $N_B(0) = 10$ (Fig.~\ref{sd_WF_nu_0p01_M_100_N_90_up_label}). We also see that the fluctuations of the molecule number around its mean,
$\mu_{1A}(t) \pm \sigma(t)$ (light blue dotted lines), decrease considerably
with increasing $N_B(0)$.% $ = M - N$, from $N_B(0) = 0$ to $N_B(0) = 10$.

For the simple autocatalytic reaction~(\ref{2nd_reaction_WF}), notable
differences between stochastic and deterministic kinetics— referred to
as “stochastic delay”—have been reported previously \cite{schuster2019special}.
Reference~\cite{schuster2019special} highlighted that these differences
are considerably larger for the autocatalytic reaction $\mathrm{A} + \mathrm{B} \rightarrow 2\mathrm{B}$ with $M = 100$ and $N_B(0) = 1$  (i.e., for $N = M - 1 = 99$) than for the simple non-autocatalytic bimolecular reaction $\mathrm{A} + \mathrm{B} \rightarrow 2\mathrm{C}$ with
$N_A(0) + N_B(0) = M = 100$ and $N_A(0) = N_B(0) = 50$. For the latter reaction, $M = 100$ is sufficiently large that stochastic and deterministic kinetics are in close agreement, but simple autocatalysis can
still exhibit substantial deviations. 

This observation motivated our decision to analyze the case $M = 100$ as well. We adopt the convention of having 100 molecules in a unit volume ($V=1$). For $V=1$, the numerical values of the ``deterministic'' (macroscopic) rate constants $\mathcal{K}_{1}$ and $\mathcal{K}_{2}$ in the kinetic rate equations [Eqs.~(\ref{Det_Chem_Kin_WF_system_A_main})–(\ref{Det_Chem_Kin_WF_only_a_solution_of})] coincide with the ``mesoscopic'' rate constants $k_1$ and $k_2$ of the CME. For $V>1$, which is considered below, we assume that $\mathcal{K}_{1}$ and $\mathcal{K}_{2}$ are volume-independent and that
\begin{equation}
\label{relationship_K_k_rate_constants}
k_1 = \mathcal{K}_{1}, \qquad k_2 = \frac{\mathcal{K}_{2}}{V}.
\end{equation}

It is worth noting that the scenario studied in Ref.~\cite{schuster2019special} closely resembles that shown in our Fig.~\ref{sd_WF_nu_0p01_M_100_N_99_up_label}, differing mainly by convention: we plot the average number and concentration of A molecules, instead those of B. This is because when $N < M$, the autocatalytic reaction (\ref{2nd_reaction_WF}) can be regarded as the limiting case of the Finke-Watkzy model (\ref{1st_reaction_WF})–(\ref{2nd_reaction_WF}) as $k_1 \to 0$. For sufficiently small $k_1/k_2$, the solutions of the CME for the FWM and for the pure autocatalytic reaction become practically indistinguishable, and the same applies to the deterministic solutions. Note that the condition $N < M$ is necessary for pure autocatalysis since the case $N = M$ (Fig.~\ref{sd_WF_nu_0p01_M_100_N_100_up_label}) results in a trivial solution where the reaction does not proceed.

Note also that we do not provide any quantitative measure of the “stochastic delay”. Various such measures have been proposed (see Ref.~\cite{schuster2019special}), but none appears clearly superior to the others. Instead, we prefer to present the plots of $a(t)$ and $a_s(t)$, as they are simpler and more informative.

\begin{figure}[h]
\begin{center}					  				
\rotatebox{270}{\scalebox{0.34}{\includegraphics{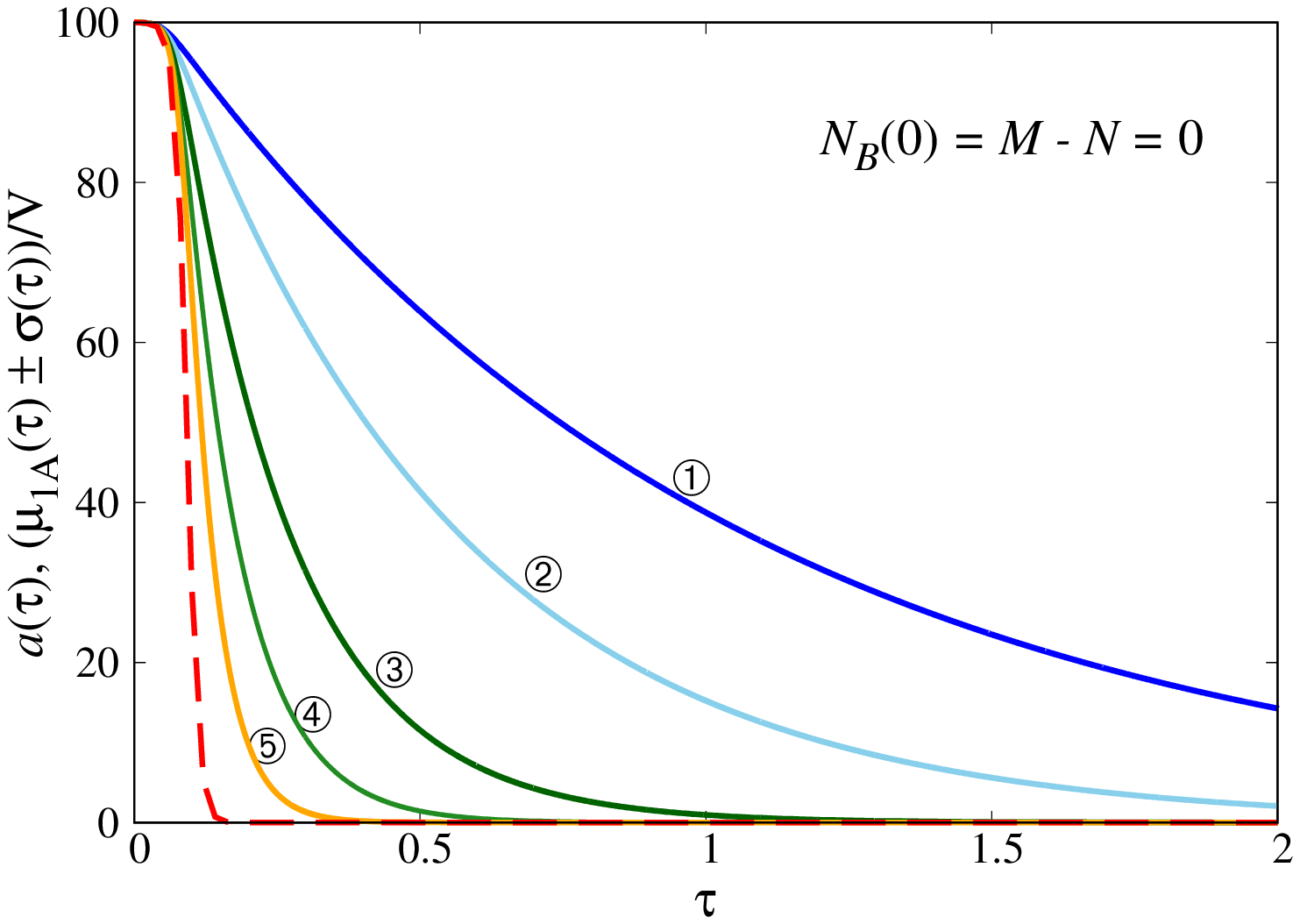}}} 
\end{center}  
\caption{Stochastic versus deterministic kinetics for the reactions (\ref{1st_reaction_WF})--(\ref{2nd_reaction_WF}) for several values of the total number of molecules $M$, with system volume scaling as $V \propto M$, initial condition $N_B(0) = 0$ ($b_0 = 0$), and macroscopic rate constants $\mathcal{K}_{1} = 0.001$ and $\mathcal{K}_{2} = 0.1$. Consequently, $k_1 = 0.001$ in all cases, while $k_2 = \mathcal{K}_{2}/V$ varies with $M$. The deterministic concentration of A, $a(t)$ (\ref{Det_Chem_Kin_WF_only_a_solution_of}), is shown as a red dashed line, and its stochastic counterpart, $a_s(t) = \mu_{1A}(t)/V$ (\ref{stochastic_concentration}), is plotted as thick solid lines for $M = 100$ (blue, 1), $200$ (light blue, 2), $500$ (dark green, 3), $1000$ (light green, 4), and $2000$ (orange, 5), ordered from left to right. The corresponding system volumes are $V = 1$, $2$, $5$, $10$, and $20$, respectively. For clarity, the standard-deviation envelope around the mean is omitted. All quantities are plotted as functions of the rescaled time variable $\tau \equiv \mathcal{K}_{2}t$.}
\label{moments_WF_nu_0p01_M_vary_N_equal_M_up_label}
\end{figure}

A systematic study of the differences between stochastic and deterministic time evolution as a function of model parameters is beyond the scope of the present work. Nevertheless, we have to address the following question: how does the stochastic delay depend on the total number of molecules in the system, $M$, assuming that the system volume scales proportionally, $V \propto M$? One expects that the relative differences between stochastic and deterministic time evolution become less pronounced as $M$ increases, even in the case most favorable to stochastic delay ($M = N$), in agreement with the general mathematical results concerning the asymptotic equivalence of the CME solution and the deterministic kinetics in the limit $V \to \infty$ \cite{kurtz1972relationship}. This expectation is indeed confirmed. In Figure \ref{moments_WF_nu_0p01_M_vary_N_equal_M_up_label} we plot the deterministic concentration $a(t)$ against $a_s(t) = \mu_{1A}(t)/V$ (\ref{stochastic_concentration}), i.e., the normalized mean number of molecules, for $M = 100$, $200$, $500$, $1000$, and $2000$ ($V = 1$, $2$, $5$, $10$, and $20$, respectively).  

In all cases we assume the same values of the macroscopic rate constants, $\mathcal{K}_{1} = 0.001$ and $\mathcal{K}_{2} = 0.1$, which implies different values of the mesoscopic rate constant $k_{2}$ according to Eq.~(\ref{relationship_K_k_rate_constants}). However, there is a unique deterministic concentration $a(t)$ with $a(0) = 100$ to be compared with. In all depicted cases, the stochastic delay remains visible, although it decreases with increasing $M$.

Even for the largest system size analyzed ($M = 2 \cdot 10^3$), the stochastic delay in the FWM is still more pronounced than in the case of pure autocatalysis (\ref{2nd_reaction_WF}) with $M = 10^2$ molecules. This demonstrates that, for certain parameter choices, substantial discrepancies between stochastic and deterministic time evolution can persist even in relatively large systems.

\begin{figure}[h]
\begin{center}					  				
\rotatebox{270}{\scalebox{0.34}{\includegraphics{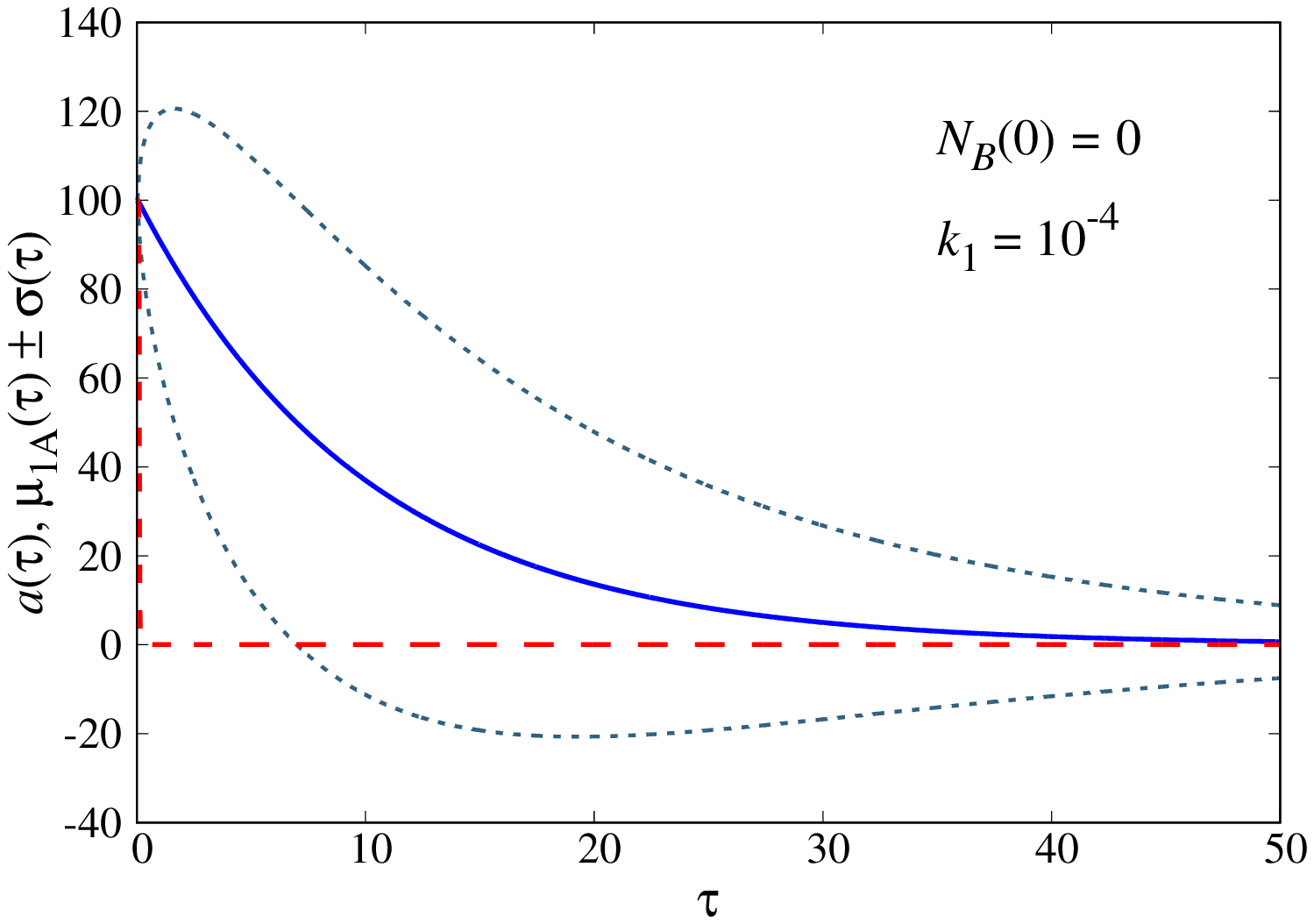}}} 
\end{center}  
\caption{Stochastic versus deterministic kinetics for reactions~(\ref{1st_reaction_WF})--(\ref{2nd_reaction_WF}), with the initial condition $N_B(0) = 0$ and rate constants $k_1 = 10^{-4}$ (ten times smaller than in all other plots) and $k_2 = 10^{-1}$. The deterministic concentration of A, $a(t)$ [Eq.~(\ref{Det_Chem_Kin_WF_only_a_solution_of})], is shown as a red dashed line and decays much faster than its stochastic counterpart, $a_s(t) = \mu_{1A}(t)/V$ [Eq.~(\ref{stochastic_concentration})], plotted as a thick blue solid line. Note the difference in the time scale compared with Fig.~\ref{sd_WF_nu_0p01_M_100_N_100_up_label}. The light blue dotted lines indicate the single–standard deviation envelope around the mean, $\mu_{1A}(t) \pm \sigma(t)$, providing a quantitative measure of stochastic fluctuations. The system volume is set to $V = 1$. All quantities are plotted as functions of the dimensionless time variable $\tau = k_2 t$.}
\label{sd_WF_nu_0p01_M_100_N_100_up_label_k1_10m4}
\end{figure}
%%---------------END OF FIGURE------------------%
%
There is yet another issue that must be addressed here—the dependence of the stochastic delay on $k_1$ at a fixed value of $k_2$ in the case $N_B = 0$, for which this effect is expected to be most pronounced for a given $k_1$.

If $k_{1} \gg k_{2}$, the dynamics are dominated by reaction~(\ref{1st_reaction_WF}). For this first-order process, however, the "stochastic" concentration $a_s(t)= \mu_{1A}(t)/V$ [Eq.~(\ref{stochastic_concentration})] follows exactly the same exponential decay law ($\sim \exp(-k_{1}t)$) as the corresponding "deterministic" concentration $a(t)$ \cite{mcquarrie1967stochastic}. Thus, although large fluctuations may occur when the total molecule number $M$ is small, no systematic differences are expected between the stochastic and deterministic time evolution of the system. As $k_1$ decreases, the discrepancies between the stochastic and deterministic descriptions grow. They become noticeable for $k_1 = 10^{-2}$ (not shown), substantial for $k_1 = 10^{-3}$ (Fig.~\ref{sd_WF_nu_0p01_M_100_N_100_up_label}), and even more pronounced for $k_1 = 10^{-4}$ (Fig.~\ref{sd_WF_nu_0p01_M_100_N_100_up_label_k1_10m4}). In the latter case, a comparison with Fig.~1 shows that the time scale of the delay increases approximately in proportion to $1/k_1$.

\subsubsection{Qualitative explanation of the observed effects}

After presenting the results, we now turn to an explanation of the observed behavior of the system defined by (\ref{1st_reaction_WF})--(\ref{2nd_reaction_WF}). We begin with an intuitive account of both the stochastic delay and the large fluctuations around the mean observed in certain cases.  

When the network (\ref{1st_reaction_WF})--(\ref{2nd_reaction_WF}) is initialized with $N_B(0)=0$, the waiting time for the first $\mathrm{A}\rightarrow\mathrm{B}$ event may be long. Because this waiting time is exponentially distributed, many realizations remain in the initial state for a substantial period before any reaction occurs, whereas deterministic kinetics proceeds without such a delay. Consequently, the ensemble mean exhibits a systematic lag that can be captured only by solving the full CME or by stochastic simulation. The same reasoning can be applied to the case $N_B(0)=1$, although here it is much less compelling.  

The explanation of the strong dependence of the CME solution on the initial number of $B$ molecules, $N_B(0)$ (Figures~\ref{sd_WF_nu_0p01_M_100_N_100_up_label}--\ref{sd_WF_nu_0p01_M_100_N_99_up_label}), is similar. For $N_B(0)=0$ ($N=M$), the transition rate $r_N$ from the initial state $n=N$ to $n=N-1$ is much smaller than in the case $N_B(0)=1$ ($N=M-1$). Consequently, the average waiting time for this first transition is significantly longer when $N_B(0)=0$.

The origin of the large fluctuations around the mean for small \(N_B(0)\) is likewise intuitive. Although the total molecule number is relatively large (\(M=100\)), the \(B\)-subsystem is initially very small (in the extreme case, absent). As a result, \(N_B(t)\) shows large relative fluctuations; by the constraint (\ref{constraint_on_mass_conservation}), the same holds for \(N_A(t)\). While suggestive, these arguments are qualitative and incomplete, and may be misleading. A more careful analysis is therefore required.

A full explanation of the discrepancies between deterministic and stochastic kinetics in the FWM (\ref{1st_reaction_WF})–(\ref{2nd_reaction_WF}) and related autocatalytic networks is beyond the scope of this work. Here, we focus instead on illustrating and comparing the exact stochastic and deterministic solutions for this particular reaction network, for which an explicit analytical solution of the CME can be obtained. A partial understanding of the origin of the stochastic delay can, however, be obtained from an analysis of the moment equations (see the next subsection).

\subsubsection{Moment equations provide partial explanation of large differences between deterministic and stochastic dynamics \label{method_of_moments}}

To study the kinetics of the Finke–Watzky model, one may replace the full CME analysis with an approximate approach based on a truncated system of moment equations. Here we examine such equations in order to provide a partial, semiquantitative explanation for the stochastic delay phenomenon. At the same time, our results highlight the limitations of moment-equation approaches when it comes to accurate quantitative prediction of the magnitude of differences between stochastic and deterministic kinetics.

In the simplest case, we consider two equations: one for the first moment and one for the second moment of the probability distribution, or equivalently for the first and second cumulant, $\kappa_{1A}(t) = \mu_{1A}(t)$ and $\kappa_{2A}(t) = \mu_{2A}(t) - \mu_{1A}^2(t)$, 
\begin{eqnarray}
\label{moment_equations_1}
\dot{\mu}_{1A}  &=& - ({k}_{1} + {k}_{2} M)\mu_{1A} + {k}_{2} \mu^2_{1A} + {k}_{2} \kappa_{2A}, \\
\label{moment_equations_2}
\dot{\kappa}_{2A} &=&  ({k}_{1} + {k}_{2} M) \mu_{1A} - {k}_{2} \mu^2_{1A} + 2 {k}_{2} \kappa_{3A} \nonumber \\
&+& \left[ 4 {k}_{2} \mu_{1A} - (2 M + 1) {k}_{2} - 2 {k}_{1} \right] \kappa_{2A},
\end{eqnarray}
where dot denotes time derivative. The closure scheme can be chosen in various ways \cite{bronstein2018variational}; the simplest is the so-called Gaussian closure, where $\kappa_{3A}(t) = 0$. Although Eqs.~(\ref{moment_equations_1})--(\ref{moment_equations_2}) appear analytically intractable regardless of the closure scheme, they still provide useful qualitative insight into the dynamics of the system at hand.

First, note that the form of Eq.~(\ref{moment_equations_1}) is independent of the chosen closure scheme or the number of moment equations considered. However, the actual values of $\kappa_{2A}(t)$, and consequently of $\mu_{1A}(t)$, do depend on the evolution equations for $\kappa_{2A}(t)$ and higher-order cumulants (if included). For $\kappa_{2A}(t)=0$, Eq.~(\ref{moment_equations_1}) reduces to the deterministic rate equation for the concentration $a(t)$ [Eq.~(\ref{Det_Chem_Kin_WF_only_a})]. Hence, the minimal nontrivial moment system consists of two equations. 

For simplicity, we now set $V=1$, so that $a_s(t)$ [Eq.~(\ref{stochastic_concentration})] reduces to $\mu_{1A}$, $a_0 = N$, $b_0 = M - N$, while the deterministic (macroscopic) rate constants $\mathcal{K}_{i}$, $i=1,2$, coincide with the mesoscopic rate constants $k_i$ of the CME. In this case, the deterministic concentration $a(t)$ evolves according to 
\begin{eqnarray}
\label{Det_Chem_Kin_WF_only_a_moments}
\dot{a} &=& - \left(k_{1} + k_{2} M \right)a + k_{2} a^2 
         = k_{2}a\left(a - \tilde{M}\right),
\end{eqnarray}
where $\tilde{M} = M + k_1/k_2$ (see Eq. (\ref{parameters_of_CME_WF_definitions})). Equations (\ref{moment_equations_1}) and (\ref{Det_Chem_Kin_WF_only_a_moments}) imply that 
\begin{eqnarray}
\label{negative_time_derivatives_of_a_and_mu_1}
\forall t \in [0,\infty): ~~~\dot{a}(t) \leq 0, ~~~ \dot{\mu}_{1A}(t) \leq 0.
\end{eqnarray}
The former inequality is immediate (as $0 \leq a \leq \tilde{M}$), while the latter follows from the relations $\tilde{M}\,\mu_{1A}(t) \geq  M\,\mu_{1A}(t) \geq N\,\mu_{1A}(t)$ and $N\,\mu_{1A}(t) \geq \mu_{2A}(t) = {\kappa}_{2A}(t) + \mu^2_{1A}(t)$.

For small values of $t$, we expand both $a(t)$ and $\mu_{1A}(t)$ in a Taylor series around $t=0$,
\begin{equation}
\label{Taylor_expansion_of_x}
x(t) = x(0) + \dot{x}(0)t + \tfrac{1}{2}\ddot{x}(0)\, t^2 + \ldots,
\end{equation}
where $x = a$ or $\mu_{1A}$. Clearly, we have
\begin{equation}
\label{initial_conditions}
a_0 = a(0) = \mu_{1A}(0) = N.
\end{equation}
If $\kappa_{2A}(0) > 0$, then Eq.~(\ref{moment_equations_1}) implies
\begin{equation}
\label{inequality_derivatives_1}
\dot{a}(0) < \dot{\mu}_{1A}(0) < 0.
\end{equation}
If, on the other hand, $\kappa_{2A}(0) = 0$ (which is the case of interest here due to the assumed initial condition~(\ref{deterministic_initial_condition})), we obtain
\begin{equation}
\label{inequality_derivatives_2}
\dot{a}(0) = \dot{\mu}_{1A}(0) = - k_2 N (\tilde{M} - N) < 0.
\end{equation}
In this situation, it is natural to assume $\dot{\kappa}_{2A}(0) > 0$ (since the probability distribution is expected to acquire nonzero variance as time evolves). The inequality $\dot{\kappa}_{2A}(0) > 0$ also follows from Eq.~(\ref{moment_equations_2}) under the assumption $\kappa_{3A}(0) = 0$, which yields
\[
\dot{\kappa}_{2A}(0) = k_2 N (\tilde{M} - N) = -\dot{\mu}_{1A}(0).
\]
Hence, provided $N \geq \lfloor \tilde{M}/2 \rfloor$, by comparing the time derivatives of Eqs.~(\ref{moment_equations_1}) and (\ref{Det_Chem_Kin_WF_only_a_moments}), we obtain
\begin{equation}
\label{inequality_derivatives_3}
\ddot{a}(0) < \ddot{\mu}_{1A}(0) = \ddot{a}(0) + k_2 \dot{\kappa}_{2A}(0),
\end{equation}
with
\[
\ddot{a}(0) = -k^2_2 N (\tilde{M} - N)(2N - \tilde{M}) < 0
\]
and
% < 0
\[
\ddot{\mu}_{1A}(0) = -k^2_2 N (\tilde{M} - N)(2N - \tilde{M} - 1) < 0.
\]
Therefore, in both cases, i.e., whether $\kappa_{2A}(0) = 0$ or $\kappa_{2A}(0) > 0$, we obtain from~(\ref{Taylor_expansion_of_x}) that at least for short times ($t \gtrsim 0$) we have
\begin{equation}
\label{inequality_values}
a(t) < \mu_{1A}(t)
\end{equation}
and, using the short-time expansion $\dot{x}(t) = \dot{x}(0) + \ddot{x}(0)\, t + \ldots$,
\begin{equation}
\label{inequality_derivatives_3A}
\dot{a}(t) < \dot{\mu}_{1A}(t).
\end{equation}
Next, consider the function
\begin{equation}
\label{g_of_x_function}
g(x) = k_2 x^2 - (k_{1} + k_{2} M)x = k_2 x (x - \tilde{M}),
\end{equation}  
which appears on the right-hand side of Eqs.~(\ref{moment_equations_1}), (\ref{moment_equations_2}) and (\ref{Det_Chem_Kin_WF_only_a_moments}). Clearly, $g(x) < 0$ for $x \in (0, \tilde{M})$. Moreover, $g(x)$ is decreasing on $[0, \tilde{M}/2)$ and increasing on $[\tilde{M}/2, \infty)$. Thus, if at at some time $t$ the condition %{inequality_derivatives_3A}
\begin{equation}
\label{first_half_of_values}
\tilde{M}/2 < a(t) < \mu_{1A}(t) < \tilde{M},
\end{equation}
is satisfied, then necessarily $g(a(t)) < g(\mu_{1A}(t))$. But since condition~(\ref{first_half_of_values}) holds for $t \gtrsim 0$ (see Eq.~(\ref{inequality_values})), and given that ${k}_{2}\kappa_{2A}(t) > 0$, from Eqs.~(\ref{moment_equations_1}) and (\ref{Det_Chem_Kin_WF_only_a_moments}) it follows that
\begin{equation}
\label{inequality_derivatives_4}
\dot{a}(t) < \dot{\mu}_{1A}(t).
\end{equation}
Therefore once the inequality~(\ref{inequality_derivatives_3A}) arises at $t\gtrsim 0$ (see Eq.~\ref{inequality_values}), it is preserved as long as condition~(\ref{first_half_of_values}) remains satisfied. The conditions (\ref{initial_conditions}) and (\ref{inequality_derivatives_4}) imply that ${a}(t) < {\mu}_{1A}(t)$, which explains the appearance of the stochastic delay as long as condition~(\ref{first_half_of_values}) is satisfied. But once $g(a) > g(\mu_{1A})$, which can happen when (\ref{first_half_of_values}) is not satisfied, inequality (\ref{inequality_derivatives_4}) may no longer hold. At a certain time $t = t_{de}$ equality is reached in Eq.~(\ref{inequality_derivatives_4}), and for $t > t_{de}$ the inequality reverses its sign. 

Nevertheless, we can still conclude that ${a}(t) < {\mu}_{1A}(t)$ for all times. Suppose that, for some $t = t_{fe} $, at which condition~(\ref{first_half_of_values}) is no longer satisfied, the curves ${a}(t)$ and ${\mu}_{1A}(t)$ intersect for the first time, i.e., ${a}(t_{fe}) = {\mu}_{1A}(t_{fe})$. Since ${a}(t) \le {\mu}_{1A}(t)$ for all $t < t_{fe}$, this would require $\dot{\mu}_{1A}(t_{fe}) < \dot{a}(t_{fe})$. However, this is impossible: a comparison of Eqs.~(\ref{moment_equations_1}) and~(\ref{Det_Chem_Kin_WF_only_a_moments}) shows that whenever ${a}(t_{fe}) = {\mu}_{1A}(t_{fe})$, one must have $\dot{\mu}_{1A}(t_{fe}) \ge \dot{a}(t_{fe})$, with equality only if $\kappa_{2A}(t_{fe})=0$. This confirms that ${a}(t) \leq {\mu}_{1A}(t)$ (and thus the stochastic delay) for all $t$.

The above analysis also reveals a close connection between the fluctuations, as quantified by the variance $\kappa_{2A}(t)=\sigma^2(t)$ or standard deviation $\sigma(t)$, and the difference between the time derivatives of $a(t)$ and $\mu_{1A}(t)$, and hence with the stochastic delay. This connection is clearly visible in our numerical results presented so far (see Figs.~\ref{sd_WF_nu_0p01_M_100_N_100_up_label}--\ref{sd_WF_nu_0p01_M_100_N_90_up_label} and Fig.~\ref{sd_WF_nu_0p01_M_100_N_100_up_label_k1_10m4}), and also follows from the form of the r.h.s. of Eq.~(\ref{moment_equations_1}). Nevertheless, it can be further elucidated by analyzing Eq.~(\ref{moment_equations_2}). 

We expect $\kappa_{2A}(t)$ to initially increase from its starting value $\kappa_{2A}(0) = 0$, and then decrease, eventually vanishing in the long-time limit: $\lim_{t \to \infty} \kappa_{2A}(t) = 0$. In the cases of greatest interest to us ($N_B(0) = 0$ and $N_B(0) = 1$), where fluctuations are expected to be moderate or large, we can show that the maximum value $\kappa_{2A}(t) = \kappa_{2A}(t_m)$ occurs at the time $t_m$ for which $\mu_{1A}(t_m) \approx M/2 \approx \tilde{M}/2$ (i.e., the reaction has reached approximately half-completion). Let us first assume that $\kappa_{3A}(t) = 0$. Writing $\kappa_{2A}(t_m)$ as $\beta^2 \tilde{M}^2$, it follows from Eq.~(\ref{moment_equations_2}) that the condition for the maximum of $\kappa_{2A}(t)$ ($\dot{\kappa}_{2A}(t_m) = 0$) can be expressed as
\begin{equation}
\label{moment_equations_2_f_g}
\mu^2_{1A} - (\tilde{M} + 4 \beta^2 \tilde{M}^2)\mu_{1A} + \beta^2 \tilde{M}^2 (2 \tilde{M} + 1) = 0.   
\end{equation}
For $N_B(0) = 0$ and even for $N_B(0) = 1$, we expect $\beta \gtrsim 0.1$, corresponding to a maximum standard deviation $\sigma$ of about $0.1$ of the total initial molecule number. However, the physically acceptable solution of the quadratic equation (Eq.~(\ref{moment_equations_2_f_g})) is a slowly increasing function of $\beta$. For instance, for ${M}=100$, $k_1 = 0.001$ and $k_2 = 0.1$  ($\tilde{M}=100.01$) and $\beta=0.1$ it equals $0.4385 \tilde{M}$, for $\beta=0.2$ it is $0.4845 \tilde{M}$, and for $\beta=0.5$ it reaches $0.4981 \tilde{M}$. Including $\kappa_{3A}(t)\neq 0$, i.e., using the full form of Eq.~(\ref{moment_equations_2}), is expected to alter the results quantitatively but only slightly. This case is effectively equivalent to solving the full CME.

\begin{figure}[h]
\begin{center}					  				
\rotatebox{270}{\scalebox{0.34}{\includegraphics{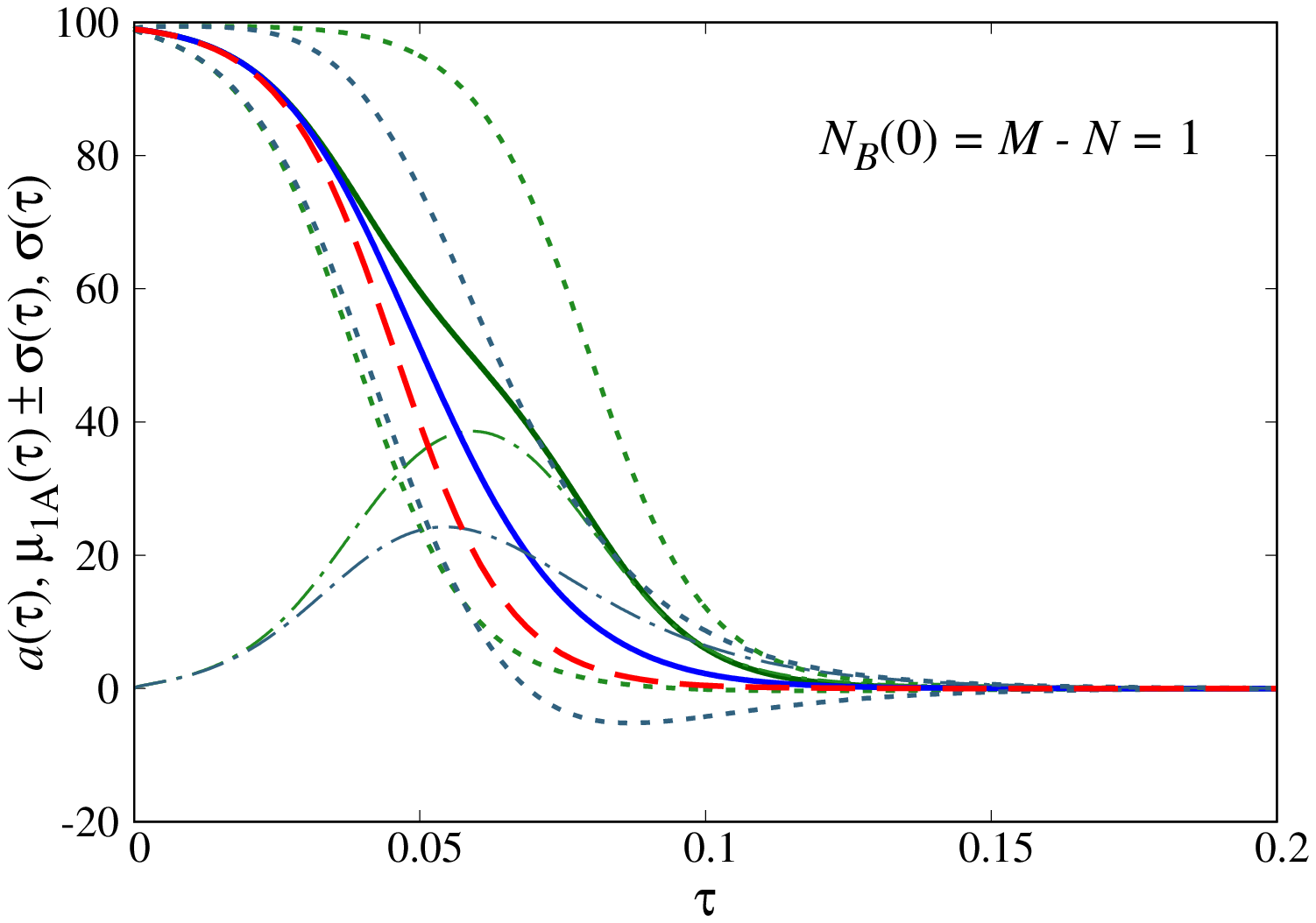}}} 
\end{center}  
\caption{Comparison of full (blue, exact solution of the CME) and approximate (green, moment equations) stochastic kinetics for reactions (\ref{1st_reaction_WF})--(\ref{2nd_reaction_WF}) with $\kappa_{3A}(t)= 0$,  $N_B(0) = 1$, $k_1 = 0.001$, and $k_2 = 0.1$. The deterministic solution $a(t)$ (\ref{Det_Chem_Kin_WF_only_a_solution_of}) is shown as a red dashed line. In each case, we plot $a_s(t) = \mu_{1A}(t)/V$ for $V=1$ (thick solid line), its standard deviation envelope $\mu_{1A}(t)\pm\sigma(t)$ (light dotted lines), and the standard deviation $\sigma(t)$ (dash-dotted lines). Considerable quantitative differences between the approximation and the exact result are evident. All quantities are shown as functions of $\tau = k_2 t$.}
\label{moments_WF_nu_0p01_M_100_N_99_up_label}
\end{figure}

All these semiquantitative predictions are borne out by the numerical results for both the CME solution and the moment-equation solution. For $\kappa_{3A}(t)= 0$ and $N_B(0)=1$, the solution of Eqs.~(\ref{moment_equations_1})--(\ref{moment_equations_2}) is shown in Fig.~\ref{moments_WF_nu_0p01_M_100_N_99_up_label}. Note that significant quantitative discrepancies are observed between this approximate solution (green) and the exact result (i.e., the solution of the CME, blue). For $N_B(0)=5$ (not shown), the approximate and exact stochastic solutions are essentially indistinguishable. However, in the most interesting case, $N_B(0)=0$, we were unable to obtain a stable numerical solution of Eqs.~(\ref{moment_equations_1})--(\ref{moment_equations_2}): despite using various methods, including those for stiff systems, and different choices of dependent variables, the solution invariably diverged.
%%---------------END OF FIGURE------------------%
%
Clearly, other truncation schemes or larger sets of moment equations can be considered. While these may yield more accurate predictions, their quality can only be assessed by comparison with the exact solution of the CME.\footnote{In fact, the FWM may serve as a valuable testbed for various moment-closure approximations \cite{bronstein2018variational}.} This provides an additional reason why the exact solution is important.

\subsection{Extension of the Finke-Watkzy model to the case of reversible reactions: $\mathrm{A} \rightleftharpoons  \mathrm{B}$, $\mathrm{A} + \mathrm{B} \rightleftharpoons  \mathrm{B} + \mathrm{B}$ \label{Results_rev}}

When the Finke–Watzky model is used to describe protein misfolding and aggregation or colloid formation in solution in the presence of a large excess of reducing agent, the assumption of irreversibility is usually well justified. However, one can also consider situations in which at least one of the reactions (\ref{1st_reaction_WF})--(\ref{2nd_reaction_WF}) is reversible. For example, if the FWM is used to model the spread of a non-fatal, curable disease, one may allow for recovery, $\mathrm{B} \rightarrow \mathrm{A}$. Even in studies of metal nanoparticle synthesis in solution, the reaction (\ref{1st_reaction_WF}) is sometimes assumed to be reversible \cite{amirjani2018modified}. In such cases, instead of (\ref{1st_reaction_WF})--(\ref{2nd_reaction_WF}), one considers the reaction set (\ref{1st_reaction_WF_rev})--(\ref{2nd_reaction_WF_rev}) with $k^{-}_{2}=0$ but $k^{-}_{1} \neq 0$. In the following, we consider this scenario as well as the case $k^{-}_{2} > 0$, which may correspond to a Lindemann-type mechanism, with $\mathrm{B}$ being the initial reactant molecule and $\mathrm{A}$ an activated reaction intermediate.\footnote{In Ref.~\cite{arslan2008kinetics}, the reversible isomerization (\ref{1st_reaction_WF_rev}) and reversible autocatalysis (\ref{2nd_reaction_WF_rev}) were treated as independent reactions and their properties compared; here, they are combined into a single reaction network.}

\subsubsection{Stochastic delay in the reaction network (\ref{1st_reaction_WF_rev})--(\ref{2nd_reaction_WF_rev})}

It is reasonable to expect that the inclusion of the inverse reactions in Eqs.~(\ref{1st_reaction_WF_rev})--(\ref{2nd_reaction_WF_rev}) does not, in general, eliminate the substantial differences between stochastic and deterministic time evolution of the system. In the following, we show that the phenomenon of ``stochastic delay'' indeed persists over a broad range of values of $k_{1}^{-}$ and $k_{2}^{-}$.

To model the stochastic kinetics of the reactions (\ref{1st_reaction_WF_rev})-(\ref{2nd_reaction_WF_rev}), instead of (\ref{CME_WF}), we use the CME (\ref{CME_WF_rev}) with 
\begin{eqnarray}
\label{r_n_of_CME_WF_rev}
r_{n} =  n[k^{+}_{1} + k^{+}_{2}(M-n)] 
\end{eqnarray}
and
\begin{eqnarray}
\label{g_n_of_CME_WF_rev}
g_{n} =  \frac{1}{2} (M-n)[2k^{-}_{1} + k^{-}_{2}(M-n-1)]. 
\end{eqnarray}
The reaction rate constants $k^{+}_{1}$ and $k^{+}_{2}$ coincide with $k^{}_{1}$ and $k^{}_{2}$ of the irreversible case (\ref{1st_reaction_WF})-(\ref{2nd_reaction_WF}). 

%%---------------END OF FIGURE------------------%
\begin{figure}[h]
\begin{center}					  				
\rotatebox{270}{\scalebox{0.34}{\includegraphics{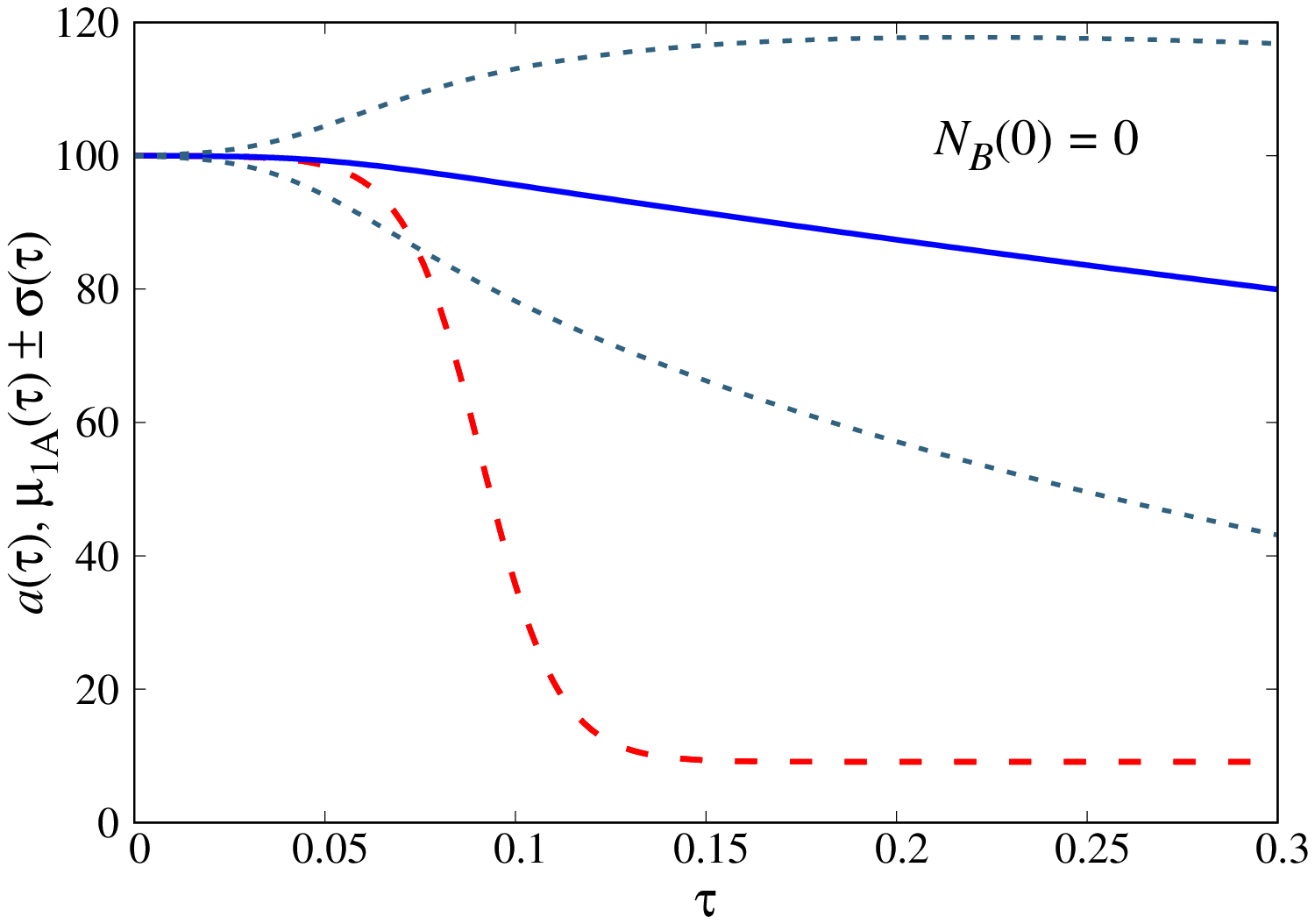}}} 
\end{center}  
\caption{Stochastic versus deterministic kinetics for the reversible Finke–Watzky model (\ref{1st_reaction_WF_rev})–(\ref{2nd_reaction_WF_rev}) with $M = N = 100$ ($N_B(0) = 0$), $k_1^{+} = k_{1}^{-} = 10^{-3}$, $k_2^{+} = 0.1$, and $k_{2}^{-} = 2 \cdot 10^{-2}$. The deterministic concentration of A molecules, $a(t)$ as given by the Eq. (\ref{Det_Chem_Kin_WF_only_a_solution_of_rev}, red dashed), decreases faster than its stochastic mean counterpart $a_s(t) \equiv \mu_{1A}(t)/V$ (\ref{stochastic_concentration}, thick blue line). Light blue dotted lines indicate the single standard deviation envelope, $\mu_{1A}(t)\pm\sigma(t)$, quantifying the magnitude of fluctuations. We set $V=1$, so that $a_s(t) = \mu_{1A}(t)$, with $\mu_{1A}(t)$ given by (\ref{first_moment_non_deg_WF}). All quantities are plotted as functions of the dimensionless time variable $\tau = k_2 t$. Although the time evolution of $a_s(t)$ and $a(t)$ differ significantly, their stationary values coincide; this is not visible in the figure because the time axis is restricted to the early stages of the time evolution to better highlight the differences in dynamics.}
\label{sd_rev_WF_nu_0p01_M_100_N_100_k_minus_0p001_0p01_label}
\end{figure}
%%---------------END OF FIGURE------------------
%%---------------END OF FIGURE------------------%
\begin{figure}[h]
\begin{center}					  				
\rotatebox{270}{\scalebox{0.34}{\includegraphics{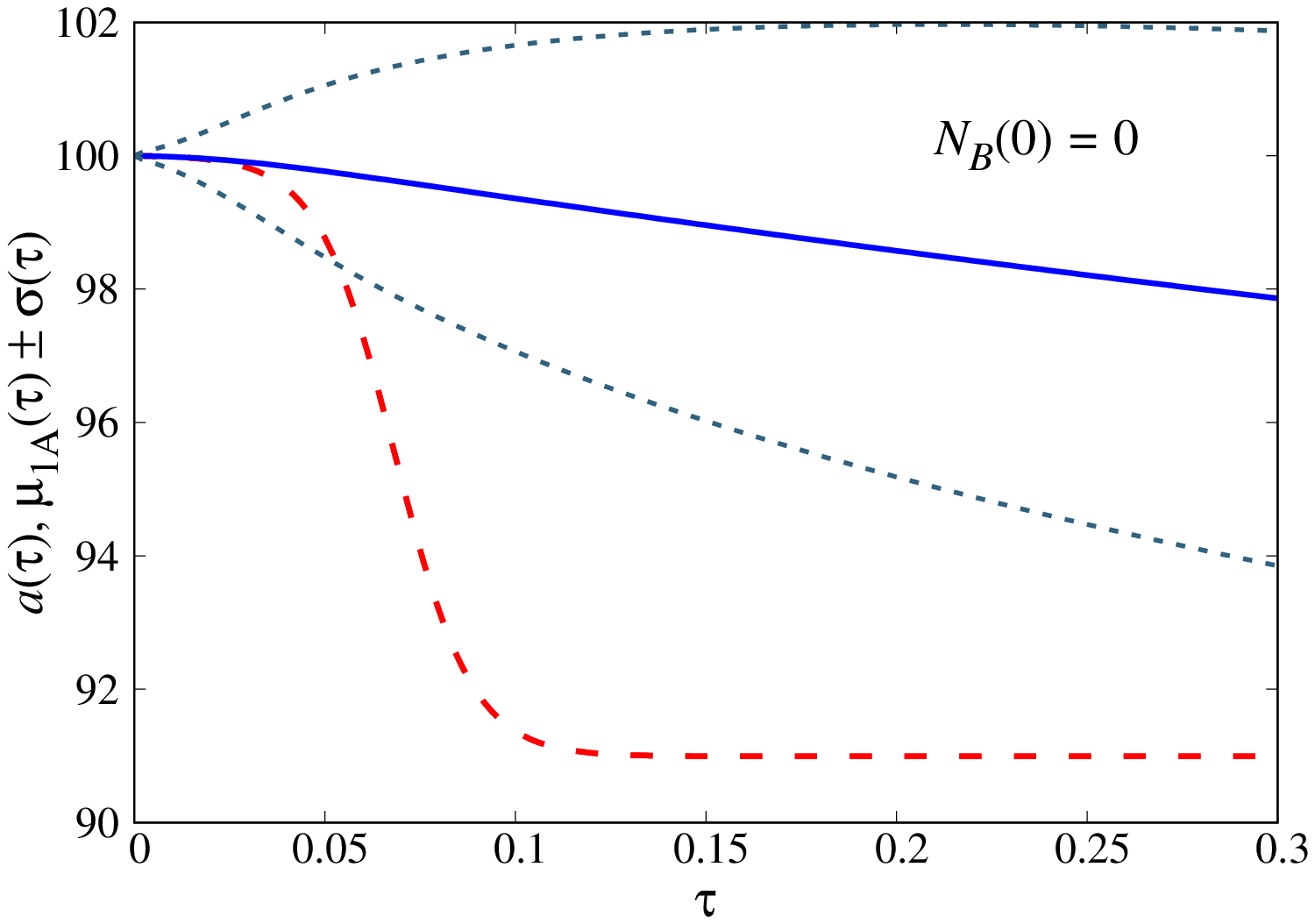}}} 
\end{center}  
\caption{Stochastic versus deterministic kinetics for the reversible Finke–Watzky model (\ref{1st_reaction_WF_rev})–(\ref{2nd_reaction_WF_rev}) with $M = N = 100$ ($N_B(0) = 0$), $k_1^{+} = 10^{-3}$, $k_2^{+} = k_{1}^{-} = 10^{-1}$, and $k_{2}^{-} = 2.0$. The deterministic concentration of A, $a(t)$ (\ref{Det_Chem_Kin_WF_only_a_solution_of_rev}, red dashed), decreases faster than the stochastic mean $a_s(t) \equiv \mu_{1A}(t)/V$ (\ref{stochastic_concentration}, thick blue). Thin lines show $\mu_{1A}(t)\pm\sigma(t)$ as a measure of fluctuations. $V=1$ so that $a_s(t) = \mu_{1A}(t)$. All quantities are plotted versus $\tau = k_2 t$. Even though, compared to Fig.~\ref{sd_rev_WF_nu_0p01_M_100_N_100_k_minus_0p001_0p01_label}, the equilibrium is shifted strongly toward the substrates, the stochastic delay remains large, as do the fluctuations around the mean number of A molecules. Despite the different dynamics of  $a(t)$ and $\mu_{1A}(t)$, the stationary values coincide; this is not visible in the figure because the time axis is restricted to better highlight the differences in dynamics.}
\label{sd_rev_WF_nu_0p01_M_100_N_100_k_minus_0p1_1_label}
\end{figure}
%%---------------END OF FIGURE------------------

For bimolecular reversible reactions, the CME (\ref{CME_WF_rev}) generally cannot be solved analytically, and numerical methods are required. This is particularly true for the reaction set (\ref{1st_reaction_WF_rev})–(\ref{2nd_reaction_WF_rev}). Here we performed direct numerical integration of the system of ODEs using routines from the GNU Scientific Library. For $M \leq 100$ this poses no difficulty. 

The corresponding deterministic rate equations, along with their solutions, are presented in Appendix~\ref{Rate_equations_rev_App}. In this case, the mapping between deterministic and stochastic rate constants reads
\begin{equation}
\label{relationship_K_k_rate_constants_rev}
k_1^{\pm} = \mathcal{K}^{\pm}_{1}, \qquad 
k_2^{+} = \frac{\mathcal{K}^{+}_{2}}{V}, \qquad 
k_2^{-} = \frac{2 \mathcal{K}^{-}_{2}}{V}.
\end{equation}

The results are shown in Figs.~\ref{sd_rev_WF_nu_0p01_M_100_N_100_k_minus_0p001_0p01_label} and \ref{sd_rev_WF_nu_0p01_M_100_N_100_k_minus_0p1_1_label} for $M = N = 100$ ($N_B(0) = 0$), $V=1$,  $k_1^{+}  = \mathcal{K}^{+}_{1}= 0.001$, and $k_2^{+} = \mathcal{K}^{+}_{2} = 0.1$ (i.e., the same values as in Fig.~\ref{sd_WF_nu_0p01_M_100_N_100_up_label}), and for two different choices of $k_{1}^{-}$ and $k_{2}^{-}$, which result in markedly different stationary solutions. In both cases, the initial condition was a 'deterministic' one, i.e., of the form given in Eq.~(\ref{deterministic_initial_condition}).

As in the irreversible case [Fig.~\ref{sd_WF_nu_0p01_M_100_N_100_up_label}], the discrepancies between stochastic and deterministic kinetics remain pronounced in both situations. This indicates that the phenomenon of ``stochastic delay'' is largely insensitive to the specific values of $k_{1}^{-}$ and $k_{2}^{-}$. What is essential is the ratio $k_1^{+}/k_2^{+}$: if this ratio is sufficiently small, pronounced differences between stochastic and deterministic kinetics are to be expected for $N_B(0) = 0$.

Note that for the chosen sets of model parameters, the stationary value of the average number of molecules normalized by the total volume, obtained either from the numerical solution of Eq.~(\ref{CME_WF_rev}) or from the analytical expression (Eq.~\ref{mu_1_hypergeom} below), is practically identical in both cases to the stationary concentration given by Eq.~(\ref{Det_Chem_Kin_WF_only_a_rev_roots}). Therefore, the difference between the stochastic and deterministic descriptions of the reaction dynamics for this system lies in the time evolution rather than in the stationary values.

\subsubsection{Steady-state solution of the CME for the reactions (\ref{1st_reaction_WF_rev})-(\ref{2nd_reaction_WF_rev})}

Although analytical expressions for $P_n(t)$ of the reaction set (\ref{1st_reaction_WF_rev})--(\ref{2nd_reaction_WF_rev}) are not accessible, the stationary distribution can be obtained straightforwardly, as is the case for other one-step reversible reactions \cite{van2007stochastic}. For completeness, we derive it below.

First, we define 
\begin{equation}
\label{CME_WF_rev_stationary_solution_WF_definitions_M}
\tilde{M} = M + \frac{k^{+}_{1}}{k^{+}_{2}}, ~~~~ \tilde{L} \equiv M + \frac{2 k^{-}_{1}}{k^{-}_{2}}, ~~~~ \theta \equiv \frac{k^{-}_{2}}{2 k^{+}_{2}}, ~~~~ \eta \equiv \frac{k^{-}_{1}}{k^{+}_{2}}.
\end{equation}

Now assume that $k^{-}_{2} \neq 0$. Using (\ref{CME_WF_rev_stationary_solution}), (\ref{r_n_of_CME_WF_rev}),   (\ref{g_n_of_CME_WF_rev}) and  (\ref{CME_WF_rev_stationary_solution_WF_definitions_M}) we get
\begin{eqnarray}
\label{CME_WF_rev_stationary_solution_WF}
P^{(s)}_n &=& \frac{1}{2^{n}}\left(\frac{k^{-}_{2}}{k^{+}_{2}}\right)^n {\binom{M}{n}}\frac{\Gamma(\tilde{L})\Gamma(\tilde{M} - n)}{\Gamma(\tilde{M})\Gamma(\tilde{L} - n)}P^{(s)}_0  \nonumber \\ &=& \theta^n {\binom{M}{n}}\frac{(1 - \tilde{L})_n}{(1 - \tilde{M})_n} P^{(s)}_0. 
\end{eqnarray}
In the above formula,  $(u)_m \equiv \Gamma(u+m)/\Gamma(u)$  denotes the Pochhammer symbol (the rising factorial). 
So far we have not needed to use the generating-function technique. However, to obtain compact analytical formulas for the first and second moments of the stationary probability distribution (\ref{CME_WF_rev_stationary_solution_WF}), we must resort to this method.\footnote{We have not been able to find compact formulas for the moments using only identities for the Gamma function and the Pochhammer symbol.} The generating function \cite{mcquarrie1967stochastic} is defined here as
\[
G(x) \equiv \sum_{n=0}^M P^{(s)}_n x^n.
\]
Once the analytical form of \(G(x)\) is known, the moments of \(P^{(s)}_n\) can be obtained by evaluating derivatives of \(G(x)\) at \(x=1\),
\begin{eqnarray}
G^{\prime}(1) &=& \sum_{n=0}^M n P^{(s)}_{n} \equiv \mu^{(s)}_{1A}, \\  G^{\prime \prime}(1) &=&  \sum_{n=0}^M n(n-1) P^{(s)}_{n} \equiv \mu^{(s)}_{2A} - \mu^{(s)}_{1A},
\end{eqnarray}
where prime denotes differentiation with respect to the auxiliary variable $x$. For the probability distribution (\ref{CME_WF_rev_stationary_solution_WF}), after some algebra, we obtain 
\begin{eqnarray}
\label{G}
G(x) &\equiv & \frac{{}_{2}F_1\left(-M, 1-\tilde{L}; 1-\tilde{M}; -\theta x\right)}{{}_{2}F_1\left(-M, 1-\tilde{L}; 1-\tilde{M}; -\theta \right)},
\end{eqnarray}
where ${}_{2}F_1\left(a, b; c; z \right)$ is the hypergeometric function. As a result, we get  

\begin{eqnarray}
\label{mu_1_hypergeom}
\mu^{(s)}_{1A} &=& \frac{M (1-\tilde{L})}{(1-\tilde{M})} \frac{{}_{2}F_1\left(1-M, 2-\tilde{L}; 2-\tilde{M}; -\theta \right)}{{}_{2}F_1\left(-M, 1-\tilde{L}; 1-\tilde{M}; -\theta \right)}\theta 
\end{eqnarray}
and
\begin{widetext}
\begin{eqnarray}
\label{mu_2_hypergeom}
\mu^{(s)}_{2A} - \mu^{(s)}_{1A}  &=& \frac{M(M-1) (1-\tilde{L})(2-\tilde{L})}{(1-\tilde{M})(2-\tilde{M})}  \frac{{}_{2}F_1\left(-M+2, 3-\tilde{L}; 3-\tilde{M}; -\theta \right)}{{}_{2}F_1\left(-M, 1-\tilde{L}; 1-\tilde{M}; -\theta \right)}\theta^2
\end{eqnarray}
\end{widetext}
where we used the basic properties of the hypergeometric function to express its derivatives using the hypergeometric function with shifted parameters.

Now consider the case when $k^{-}_{2} = 0$. Then we obtain
\begin{equation}
\label{CME_WF_rev_stationary_solution_WF_bis}
\frac{P^{(s)}_n}{P^{(s)}_0} = \left(\frac{k^{-}_{1}}{k^{+}_{2}}\right)^n {\binom{M}{n}}\frac{\Gamma(\tilde{M} - n)}{\Gamma(\tilde{M})}  =  {\binom{M}{n}}\frac{\left( - \eta \right)^n}{(1-\tilde{M})_n},
\end{equation}
where $\eta$ was defined in the equation  (\ref{CME_WF_rev_stationary_solution_WF_definitions_M}). For the distribution (\ref{CME_WF_rev_stationary_solution_WF_bis}) generating function can be obtained analogously as in the $k^{-}_{2} \neq 0$ case.  We get
\begin{equation}
\label{G_bis}
G(x) =  \frac{{}_{1}F_1\left(-M; 1-\tilde{M};  \eta x\right)}{{}_{1}F_1\left(-M; 1-\tilde{M};  \eta \right)} =  \frac{L^{(-\tilde{M})}_{M}(\eta x)}{L^{(-\tilde{M})}_{M}(\eta)}  
\end{equation}
where ${}_{1}F_1\left(a, b; z \right)$ denotes the confluent  hypergeometric function while $L^{(\alpha)}_{M}(z)$ is generalized Laguerre polynomial of degree $M$ \cite{abramowitz1968handbook}. Using the properties of the generalized Laguerre polynomials, we obtain
\begin{equation}
\label{G_bis_average1}
\mu^{(s)}_{1A} = \sum_{n=0}^M n P^{(s)}_{n}   = -\eta \frac{L^{(1-\tilde{M})}_{M-1}(\eta)}{L^{(-\tilde{M})}_{M}(\eta)}  
\end{equation}
and
\begin{equation}
\label{G_bis_average2}
\mu^{(s)}_{2A} - \mu^{(s)}_{1A} = \sum_{n=0}^M n(n-1) P^{(s)}_{n}   =  \eta^2 \frac{L^{(2-\tilde{M})}_{M-2}(\eta)}{L^{(-\tilde{M})}_{M}(\eta)}.  
\end{equation}
We see that (\ref{G}) and (\ref{G_bis}) are of the same type as the formulas for $G(x)$ derived for other simple reversible reactions \cite{mcquarrie1967stochastic} ($\mathrm{A}  + \mathrm{B} \rightleftharpoons  \mathrm{C} + \mathrm{D}$, $\mathrm{A}  + \mathrm{B} \rightleftharpoons 2 \mathrm{C}$, $2\mathrm{A} \rightleftharpoons \mathrm{C} + \mathrm{D}$, $2\mathrm{A} \rightleftharpoons \mathrm{C}$): they are expressed by either a hypergeometric function or a confluent hypergeometric function. 

%In practice, it may be easier to compute the lowest moments of the probability distributions $\{P^{(s)}_n\}_{n=0}^M$ (\ref{CME_WF_rev_stationary_solution_WF}) and (\ref{CME_WF_rev_stationary_solution_WF_bis}) not by derivatives of generating functions, but directly from the definitions of the moments and expressions (\ref{CME_WF_rev_stationary_solution_WF}) and (\ref{CME_WF_rev_stationary_solution_WF_bis}): although numerical, the result obtained will of course be exact. 

%%%%%%%%%%%%%%%

\subsection{Two Reaction Networks Closely Related to the Finke–Watzky Model}
\label{Two_other_reactions_Appendix}

In this Section, we consider two reaction sets closely related to the original Finke–Watzky model (FWM). For both of them, an explicit analytical solution of the CME can be obtained using the results already derived for the FWM (\ref{1st_reaction_WF})-(\ref{2nd_reaction_WF}).

The first, which we call the inverse Finke–Watzky model, arises by setting $k^{+}_{1}=k^{+}_{2}=0$ in (\ref{1st_reaction_WF_rev})–(\ref{2nd_reaction_WF_rev}), leaving only the reverse reactions. If we consider these reverse reactions in isolation and swap the labels A $\leftrightarrow$ B, we obtain
\begin{align}
\label{1st_reaction_inv_WF}
\mathrm{A} &\xrightarrow{k_1} \mathrm{B}, \\
\label{2nd_reaction_inv_WF}
\mathrm{A} + \mathrm{A} &\xrightarrow{k_2} \mathrm{A} + \mathrm{B}.
\end{align}
This reaction set describes, for example, a population in which individuals (e.g., duelists or predators) may die spontaneously at rate $k_1$ and, upon encounter, one of two individuals may die with probability determined by $k_2$. While a direct chemical analogue for (\ref{1st_reaction_inv_WF})–(\ref{2nd_reaction_inv_WF}) is less obvious, the bimolecular step (\ref{2nd_reaction_inv_WF}) appears in the Robertson model \cite{robertson1966numerical}, a classic example of a stiff ODE system. Irrespective of interpretation, this network provides a nontrivial example whose CME admits an explicit analytical solution.

Moreover, the inverse Finke–Watzky model provides a simple example of a reaction network for which stochastic delay is absent and whose behavior is similar to that of the simple bimolecular reaction $\mathrm{A} + \mathrm{B} \rightarrow \mathrm{C}$. Therefore, this model provides a natural reference point against which the original FWM (\ref{1st_reaction_WF})–(\ref{2nd_reaction_WF}) can be compared.

Using our results for the FWM, we can likewise obtain the explicit analytical solution of the CME for a set of two consecutive irreversible reactions—pairing the “inverse autocatalysis” (\ref{2nd_reaction_inv_WF})  with the autocatalytic step (\ref{2nd_reaction_WF}):
\begin{align}
\label{1st_reaction_consecutive}
\mathrm{A} + \mathrm{A} &\xrightarrow{k_a} \mathrm{A} + \mathrm{B}, \\
\label{2nd_reaction_consecutive}
\mathrm{A} + \mathrm{B} &\xrightarrow{k_b} \mathrm{B} + \mathrm{B}.
\end{align}
Depending on the values of the rate constants $k_a$ and $k_b$, the reaction network (\ref{1st_reaction_consecutive})–(\ref{2nd_reaction_consecutive}) can belong either to the same class as the FWM (\ref{1st_reaction_WF})–(\ref{2nd_reaction_WF}) or to class of the inverse FWM (\ref{1st_reaction_inv_WF})–(\ref{2nd_reaction_inv_WF}). Thus, for a suitable choice of model parameters, one expects the stochastic kinetics of (\ref{1st_reaction_consecutive})–(\ref{2nd_reaction_consecutive}) to exhibit both stochastic delay and sensitivity to the initial number of B molecules, as observed for the FWM.

\subsubsection{The inverse Finke–Watzky model: \(\mathrm{A} \rightarrow \mathrm{B}\), \(\mathrm{A} + \mathrm{A} \rightarrow \mathrm{A} + \mathrm{B}\)}
\label{Results_inv}

We begin with (\ref{1st_reaction_inv_WF})–(\ref{2nd_reaction_inv_WF}), for which the CME coefficients \(r_{n}\) take the form
\begin{equation}
\label{r_ns_of_CME_inv_WF}
r_{n} = \tfrac{1}{2} k_2\,n\,(n + \xi),
\end{equation}
where
\begin{equation}
\label{r_ns_of_CME_inv_WF_xi}
\xi = \frac{2 k_1}{k_2} - 1.
\end{equation}

The form of \(r_n\) in (\ref{r_ns_of_CME_inv_WF}) is very similar to that for the irreversible bimolecular reaction \(\mathrm{A} + \mathrm{B} \rightarrow 2\mathrm{C}\) (see Refs.~\cite{mcquarrie1967stochastic, laurenzi2000analytical}). In contrast, here \(\xi\) is not the initial number of A molecules and need not be a positive integer.

More importantly, however, the mapping
\[
k_2/2 \;\longleftrightarrow\; -k_2,\quad
\xi \;\longleftrightarrow\; -\tilde{M}
\]
transforms \(r_n\) in (\ref{r_ns_of_CME_inv_WF}) into the reaction rates \(r_n\) of the Finke–Watzky model (\ref{1st_reaction_WF})–(\ref{2nd_reaction_WF}), and vice versa. Hence, the analytical solution of the CME for (\ref{1st_reaction_inv_WF})–(\ref{2nd_reaction_inv_WF}) follows immediately from that for (\ref{1st_reaction_WF})–(\ref{2nd_reaction_WF}). As with the FWM, \(-r_n\) is the \(n\)th eigenvalue of \(\mathbb{W}\) (\ref{W_matrix_explicit_form}). For \(\xi > -1\) the spectrum is nondegenerate. At \(\xi = -1\), \(r_0 = r_1 = 0\), but \(n=0\) remains inaccessible, so degeneracy is again avoided, simplifying the solution.

Explicitely, with \(r_{n}\) (\ref{r_ns_of_CME_inv_WF}), the coefficients \(C_{nk}\) in (\ref{CME_WF_solution_n_t_main}) become
\begin{widetext}
\begin{equation}
\label{explicit_form_of_the_C_n_k_coefficients_inv_WF}
C_{nk} = (-1)^{\,k-n}\,(2k + \xi)\,\binom{N}{k}\binom{k}{n}\,
\frac{\Gamma(N+1+\xi)\,\Gamma(n+k+\xi)}
     {\Gamma(N+k+1+\xi)\,\Gamma(n+1+\xi)}.
\end{equation}
\end{widetext}
Assuming the deterministic initial condition (\ref{deterministic_initial_condition}), \(P_n(t)\) follows directly. If \(\xi = m\in\mathbb{Z}\) and $\xi \ge -1$, this reduces to
\begin{equation}
\label{explicit_form_of_the_C_n_k_coefficients_inv_WF_k1_0}
C_{nk}
=(-1)^{\,k-n}\,
\frac{2k + m}{n + k + m}
\binom{N}{k}\binom{k}{n}
\frac{\binom{n+k+m}{k}}{\binom{N+k+m}{k}}.
\end{equation}
The first two moments are given by (\ref{first_moment_non_deg_WF}) and (\ref{second_moment_non_deg_WF}) but with the coefficients $E^{(1)}_l$ and $E^{(2)}_l$ now given by 
\begin{align}
E^{(1)}_l&= \frac{N!}{(N-l)!}
\frac{(2l+\xi)\,\Gamma(N+1+\xi)}
     {\Gamma(N+l+1+\xi)},\label{first_moment_non_deg_inv_WF}\\
E^{(2)}_l& =(l^2+\xi l-\xi)E^{(1)}_l.\label{second_moment_non_deg_inv_WF}
\end{align}

Note that, as in the case of $r_n$ in Eq.~(\ref{eigenvalues_of_CME_WF}), the dependence of the CME rate coefficients $r_n$ in Eq.~(\ref{r_ns_of_CME_inv_WF}) on $n$ is quadratic. However, unlike in the original FWM, the coefficient of $n^2$ is positive. This seemingly minor difference leads to a profound change in the behavior of the system.  

In the case of the FWM [Eqs.~(\ref{1st_reaction_WF})–(\ref{2nd_reaction_WF})], for \(N \ge \lfloor \tilde{M}/2 \rfloor\), the sequence \(\{r_n\}\) is non-monotonic: it increases for \(n \le \lfloor \tilde{M}/2 \rfloor\) (or \(n \le \lfloor \tilde{M}/2 \rfloor + 1\)) and then decreases, assuming $\tilde{M} \approx M$, which is the most relevant situation from our viewpoint. In particular, for \(N = M\) the difference between $r_N$ and $r_{N-1}$ can be very substantial.

By contrast, in the present case $r_n$ grows monotonically with $n$, and the solution of the CME depends only on the initial number of $A$ molecules, $N$, and not on the total number of molecules, $M$ (cf. Eqs.~(\ref{explicit_form_of_the_C_n_k_coefficients_inv_WF})–(\ref{second_moment_non_deg_inv_WF})). Consequently, the solution of the CME should not be very sensitive to the choice of initial condition: the plots of $a_s(t) = \mu_{1A}(t)/V$ for $N_B(0) = 0$ and $N_B(0) = 1$ are expected to be nearly identical.  

In contrast to the FWM, we also do not expect pronounced deviations between stochastic and deterministic kinetics. This prediction is reinforced by the mathematical similarity of the CME corresponding to the reaction network Eqs.~(\ref{1st_reaction_inv_WF})–(\ref{2nd_reaction_inv_WF}) to that of the simple bimolecular reaction $\mathrm{A} + \mathrm{B} \rightleftharpoons 2 \mathrm{C}$, for which no stochastic delay is observed.  
\begin{figure}[h]
\begin{center}					  				
\rotatebox{270}{\scalebox{0.34}{\includegraphics{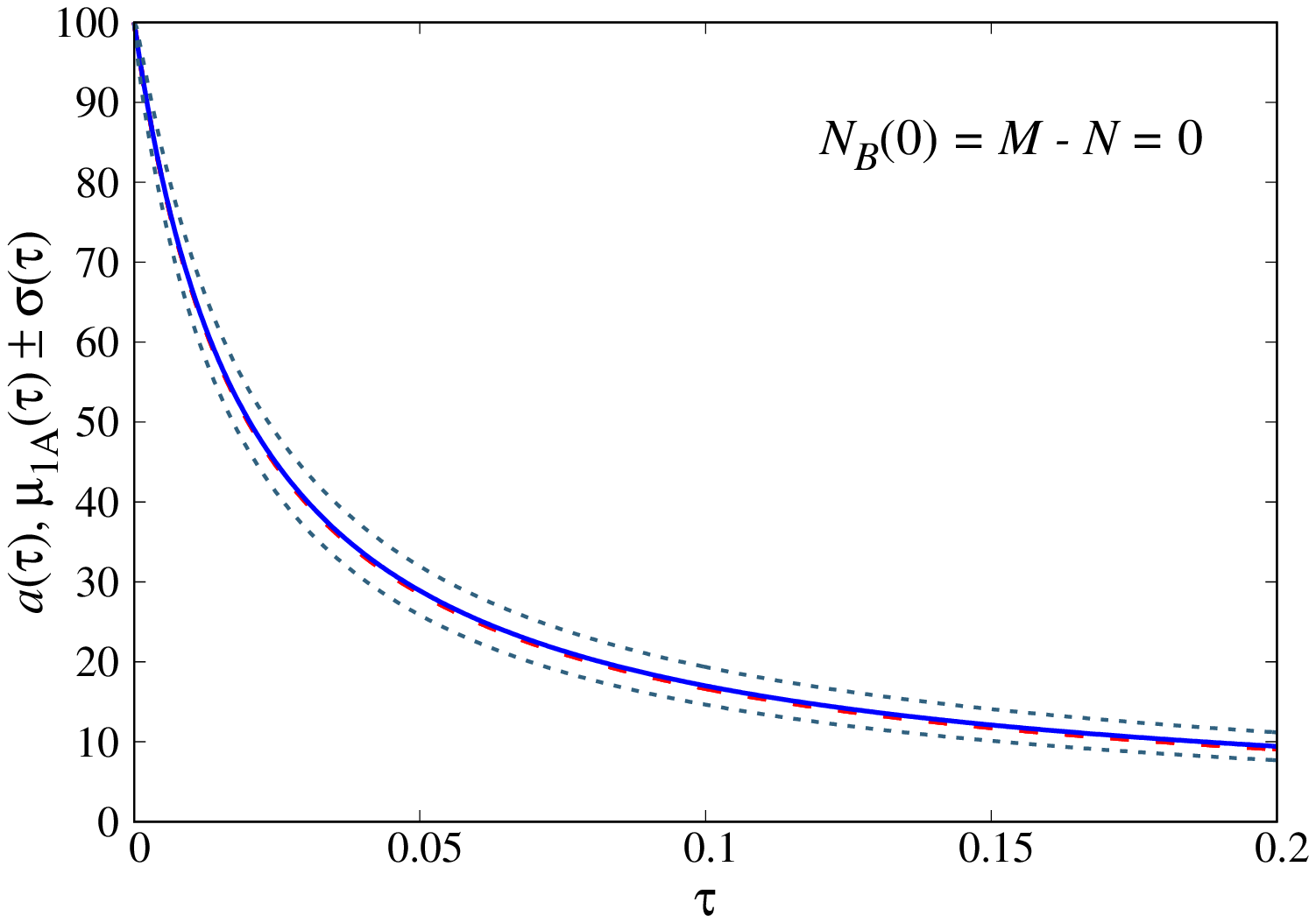}}} 
\end{center}  
\caption{Stochastic versus deterministic kinetics for the reactions 
(\ref{1st_reaction_inv_WF})–(\ref{2nd_reaction_inv_WF}), with the initial 
condition $N_B(0) = 0$ and rate constants $k_1 = 0.001$ and $k_2 = 0.1$. 
The deterministic concentration $a(t)$, given by 
Eq.~(\ref{Det_Chem_Kin_inv_WF_only_a_solution_of}), is plotted as a red 
dashed line. Its stochastic counterpart, 
$a_s(t) = \mu_{1A}(t)/V$ [Eq.~(\ref{stochastic_concentration})], 
is shown as a thick blue solid line. No visible difference is observed 
between the two time evolutions. Light blue dotted lines represent the single–standard deviation envelope around the mean, $\mu_{1A}(t) \pm \sigma(t)$, thus quantifying stochastic fluctuations, which are now much smaller than in the case of the FWM (\ref{1st_reaction_WF})--(\ref{2nd_reaction_WF}) for the same model parameters. The system volume is set to $V = 1$, so that $k_2 = 2\mathcal{K}_2$ (see Eq. (\ref{relationship_K_k_rate_constants_inv})) and 
$a_s(t) = \mu_{1A}(t)$ with $\mu_{1A}(t)$ given by 
Eq.~(\ref{first_moment_non_deg_WF}). All quantities are plotted as 
functions of the dimensionless time variable $\tau = k_2 t$.}
\label{moments_INV_WF_nu_0p01_M_100_N_100}
\end{figure}

The above predictions are confirmed by numerical results. First, no stochastic delay is observed, as illustrated in Fig.~\ref{moments_INV_WF_nu_0p01_M_100_N_100}. In the present case, the deterministic rate equation for $a(t)$ is given by Eq.~(\ref{Det_Chem_Kin_inv_WF_only_a_solution_of}).  

Second, unlike in the case of the FWM, the dependence of the solution on the initial number of $B$ molecules is only weak: the time evolution of $a_s(t) = \mu_{1A}(t)/V$ for $N_B(0) = 0$ and $N_B(0) = 1$ (not shown) is very similar.  

Finally, in contrast to the original FWM, where the second time derivative of both $a(t)$ and $\mu_{1A}(t)$ changes sign, here the plots of $a(t)$ and $\mu_{1A}(t)$ remain convex.  

Note also that the numerical evaluation of the coefficients $E^{(1)}_l$ [Eq.~(\ref{first_moment_non_deg_inv_WF})] and $E^{(2)}_l$ [Eq.~(\ref{second_moment_non_deg_inv_WF})] is considerably simpler in the present case than in the original FWM. All coefficients are positive, and for $M=100$ and $N=100$ their magnitudes do not exceed $10^{1}$ (for $E^{(1)}_l$) and $10^{3}$ (for $E^{(2)}_l$). As a result, arbitrary-precision arithmetic is not required for the numerical calculations.

\subsubsection{Two consecutive bimolecular reactions: \(\mathrm{A} + \mathrm{A} \rightarrow \mathrm{A} + \mathrm{B}\), \(\mathrm{A} + \mathrm{B} \rightarrow 2\mathrm{B}\)}
\label{Results_two_consecutive}

The last model we consider consists of the two consecutive reactions (\ref{1st_reaction_consecutive})–(\ref{2nd_reaction_consecutive}). In this case, the CME coefficients \(r_n\) are
\begin{equation}
\label{r_ns_of_CME_consecutive}
r_{n} = \tfrac12\,k_a\,n\bigl[(1-\chi)\,n + M\chi - 1\bigr],
\end{equation}
where
\begin{equation}
\chi = \frac{2k_b}{k_a},
\end{equation}
and \(M\) is the total (constant) number of molecules.

From a mathematical perspective, the solution of the CME for each of the three models considered here can be obtained from that of the others by an appropriate mapping of model parameters. However, from the viewpoint of properties relevant to chemistry or population dynamics it is appropriate to treat the original Finke–Watzky model (\ref{1st_reaction_WF})–(\ref{2nd_reaction_WF}) and the inverse Finke–Watzky model (\ref{1st_reaction_inv_WF})–(\ref{2nd_reaction_inv_WF}) on an equal footing: they belong to distinct classes, one exhibiting pronounced stochastic delay and the other not. The reaction network (\ref{1st_reaction_consecutive})–(\ref{2nd_reaction_consecutive}), on the other hand, belongs, depending on the choice of model parameters, either to the same class as (\ref{1st_reaction_WF})–(\ref{2nd_reaction_WF}) or to that of (\ref{1st_reaction_inv_WF})–(\ref{2nd_reaction_inv_WF}).

Therefore, it is convenient to view the solution for (\ref{1st_reaction_consecutive})--(\ref{2nd_reaction_consecutive}) as derived either from the solution of the FWM [Eqs.~(\ref{1st_reaction_WF})--(\ref{2nd_reaction_WF})] or from the solution of the inverse FWM [Eqs.~(\ref{1st_reaction_inv_WF})--(\ref{2nd_reaction_inv_WF})], depending on $\chi$ and $M$.

If \(\chi>1\), then \(M\chi>1\) and (\ref{r_ns_of_CME_consecutive}) becomes
\begin{equation}
\label{r_ns_of_CME_consecutive_ii}
r_{n} = \tfrac12\,k_a(\chi-1)\,n\Bigl[\frac{M\chi-1}{\chi-1}-n\Bigr],
\end{equation}
which matches the form of (\ref{eigenvalues_of_CME_WF}) under the identification
\begin{equation}
\label{r_ns_of_CME_consecutive_ii_identification}
\tfrac12\,k_a(\chi-1)=k_2,\quad \frac{M\chi-1}{\chi-1}=\tilde{M}>M.
\end{equation}
Thus, for large values of $\chi$ ($k_a \ll k_b$), we have $\tilde{M} \gtrsim M$, and by analogy with the original Finke–Watzky model, a significant stochastic delay can be expected when the system is initialized with no B molecules. This is evident even without detailed analysis: in this parameter regime, reaction~(\ref{1st_reaction_consecutive}) effectively plays the role of the first-order process~(\ref{1st_reaction_WF}) in the original Finke–Watzky model, providing a low-probability “nucleation” or “seeding” event that is subsequently amplified by autocatalysis~(\ref{2nd_reaction_consecutive}).

For \(\chi=1\), \(r_n\) in (\ref{r_ns_of_CME_consecutive}) reduces to that of a pseudo–first-order reaction,
\begin{equation}
k_1=\tfrac12\,k_a(M-1),
\end{equation}
since the reaction rate becomes proportional to the number of remaining molecules \(M-1\).

Finally, if \(\chi<1\) ($2 k_b < k_a$), then
\begin{equation}
\label{r_ns_of_CME_consecutive_i}
r_{n} = \tfrac12\,k_a(1-\chi)\,n\Bigl[n+\frac{M\chi-1}{1-\chi}\Bigr],
\end{equation}
which matches (\ref{r_ns_of_CME_inv_WF}) under
\begin{equation}
\label{r_ns_of_CME_consecutive_i_identification}
k_a(1-\chi)=k_2,\quad \frac{M\chi-1}{1-\chi}=\xi.
\end{equation}
It is evident that in the limit  $k_a \gg k_b$ , the behavior of the system described by Eqs. (\ref{1st_reaction_consecutive})–(\ref{2nd_reaction_consecutive}) is predominantly governed by the first of these two reactions.

An analogous mapping between parameters of (\ref{1st_reaction_consecutive})–(\ref{2nd_reaction_consecutive}) and those of the FWM (\ref{1st_reaction_WF})–(\ref{2nd_reaction_WF}) or the inverse FWM (\ref{1st_reaction_inv_WF})–(\ref{2nd_reaction_inv_WF}) also holds at the level of the deterministic rate equations.

\section{Summary and Discussion}

In this work, we examine differences between deterministic and stochastic kinetics of the Finke-Watkzy model (FWM): irreversible autocatalysis, $\mathrm{A} + \mathrm{B} \rightarrow 2\mathrm{B}$ (\ref{2nd_reaction_WF}), paired with an irreversible (pseudo)first-order reaction, $\mathrm{A} \rightarrow \mathrm{B}$ (\ref{1st_reaction_WF}). 

As we show here, depending on the model parameters, the deviations between the time evolution of the average number of molecules (given by the solution of the Chemical Master Equation, CME) and the corresponding time dependence of the concentration (given by the deterministic rate equations) can be exceptionally large for this reaction set.

The relatively large difference between the time dependence of the average number of molecules and the time dependence of the concentration of the same chemical species has already been reported for the autocatalytic reaction (\ref{2nd_reaction_WF}) and termed 'stochastic delay' \cite{schuster2019special}. By 'relatively large' we mean large compared to a simple non-autocatalytic irreversible bimolecular reaction $\mathrm{A}  + \mathrm{B} \rightarrow 2 \mathrm{C}$ for a system with the same total number of molecules. Here we show that the presence of the second, parallel first-order reaction can greatly amplify the discrepancies between stochastic and deterministic kinetics. This is the case when there are initially no B molecules and the rate constant $k_1$ for the first-order reaction (\ref{1st_reaction_WF}) is much smaller than the rate constant $k_2$ for the autocatalytic reaction (\ref{2nd_reaction_WF}). The fluctuations around the average number of A molecules are also very large in this case. 

However, if we initially add just a single B molecule to the system, the time evolution becomes quantitatively very different. This suggests that approximate stochastic methods, in which the molecule number is treated as a continuous variable—such as the chemical Langevin equation or the chemical Fokker–Planck equation \cite{gillespie2007stochastic}—are of little use for studying the time evolution of this system. In addition, because we are interested in the regime $k_1 \ll k_2$ characterized by a strong separation of reaction time scales, the standard SSA becomes highly inefficient. Furthermore, SSA variants designed to cope with this difficulty are not equivalent to the CME, as they no longer generate exact samples from the true distribution.

Therefore, to investigate the stochastic delay effect in the Finke–Watzky model, we derive the exact analytical solution of its CME. We obtain not only explicit analytical formulas for the time dependence of the probability distribution, but also for the time evolution of its two lowest moments.

The analytical solution of the CME for the FWM allows us to find, with very little effort, solutions of the CME for two other related chemical reaction networks. These are the remaining two out of the three different reaction pairs that can be formed from the first-order decay $\mathrm{A} \rightarrow \mathrm{B}$ (\ref{1st_reaction_WF}), autocatalysis $\mathrm{A} + \mathrm{B} \rightarrow 2\mathrm{B}$ (\ref{2nd_reaction_WF}) and 'inverse autocatalysis' or 'partial binary annihilation' $\mathrm{A} + \mathrm{A} \rightarrow \mathrm{A} + \mathrm{B}$ (\ref{2nd_reaction_WF_rev}).

We have also analyzed a generalization of the FWM to reversible reactions (\ref{1st_reaction_WF_rev})–(\ref{2nd_reaction_WF_rev}), for which analytical solutions of the corresponding CME are not yet available. Nevertheless, numerical results show that the presence of inverse reactions does not eliminate the significant differences between stochastic and deterministic kinetics.

This work represents a generalization of the preliminary study of the stochastic delay effect \cite{schuster2019special}. Our findings further suggest that a pronounced stochastic delay—much larger than in the case of simple autocatalysis analyzed in Ref.~\cite{schuster2019special}—occurs not only in the Finke–Watzky model (\ref{1st_reaction_WF})–(\ref{2nd_reaction_WF}), but also in other reaction networks. In particular, substantial discrepancies between stochastic and deterministic kinetics are expected in networks that combine a first-order reaction, $\mathrm{A} \rightarrow \mathrm{B}$, with higher-order autocatalysis (e.g., $2\mathrm{A} + \mathrm{B} \rightarrow \mathrm{A} + 2\mathrm{B}$ or $\mathrm{A} + 2\mathrm{B} \rightarrow 3\mathrm{B}$).\footnote{For the reaction networks $\{\mathrm{A} \rightarrow \mathrm{B}, 2 \mathrm{A} + \mathrm{B} \rightarrow \mathrm{A} + 2\mathrm{B}\}$ and $\{\mathrm{A} \rightarrow \mathrm{B}, \mathrm{A} + 2\mathrm{B} \rightarrow 3\mathrm{B}\}$, this has indeed been confirmed by our preliminary numerical results.}

Our results also extend those of Refs.~\cite{laurenzi2000analytical, arslan2008kinetics}, enlarging the class of reaction networks for which the CME has been solved analytically \cite{mcquarrie1967stochastic, lee2012analytical}. They can be used to test numerical results, approximate methods, or the limits of the applicability of deterministic chemical kinetics. In cases where deterministic kinetics is not applicable to a given model, the solutions of the CMEs presented here can serve as a minimal stochastic model of the system under consideration.

\section*{Data Availability}

All raw data and source codes used to generate the figures in this manuscript are openly available in Ref.~\cite{6VCFTA_2025}.

\section*{Acknowledgments}
The authors would like to thank Aleksandra Siklitckaia, Adam Kubas, Sanat Kumar Mahankudo, Marcin Rubin, and  Gonzalo Angulo. TB was supported by the National Science Centre Grant Opus 20 Project No. 2020/39/B/ST4/01952. \\

\appendix

\begin{widetext}

\section{Laplace Transform Solution of the Chemical Master Equation (\ref{CME_WF})  \label{Solution_of_CME_WF_Appendix}}

The set of equations (\ref{CME_WF}) can be solved using the Laplace transform. When applied to both sides of (\ref{CME_WF}), it yields:
\begin{eqnarray}
\label{CME_WF_LT}
s \Theta_{n}(s) - P_{n}(0) = r_{n+1}\Theta_{n+1}(s) - r_{n} \Theta_{n}(s),
\end{eqnarray}
where 
\begin{eqnarray}
\label{Theta_s_definition}
\Theta_{n}(s)  = \mathcal{L}[P_{n}(t)](s) = \int_0^{\infty} e^{-st} P_{n}(t) dt.
\end{eqnarray}
(\ref{CME_WF_LT}) is satisfied not only for $n = 1, 2, \ldots N - 1$, but also for $n=0$ and $n=N$, due to the fact that $P_{N+1} = 0$ and $r_{0} = 0$.

We assume a deterministic initial condition (\ref{deterministic_initial_condition}), namely $P_{n}(0) = 0$ for $n = 0, 1, 2, \ldots N - 1$ and $P_{N}(0) = 1$. In such a case, the solution of (\ref{CME_WF_LT}) is
\begin{eqnarray}
\label{CME_WF_LT_solution_N}
\Theta_{N}(s) & = & \frac{1}{(s + r_{N})} 
\end{eqnarray}
and %= \frac{\prod_{j=n+1}^N  r_{j}}{\prod_{j=n}^N (s + r_{j})}
\begin{eqnarray}
\label{CME_WF_LT_solution_n}
\Theta_{n}(s) & = & \frac{r_{n+1} r_{n+2} \cdots r_{N}}{(s + r_{n})(s + r_{n+1})  \cdots (s + r_{N-1})(s + r_{N})}  = \frac{1}{s + r_{n}}  \prod_{j=n+1}^N  \frac{r_{j}}{s + r_{j}}.
\end{eqnarray}
for $n = 1, 2, \ldots N - 1$. The value of $\Theta_{0}(s)$ can be obtained either from the Laplace transform of the normalization condition ($\sum_{n=0}^N P_{n}(t) ={1}$),
\begin{eqnarray}
\label{CME_WF_LT_solution_nalt}
\sum_{n=0}^N \Theta_{n}(s) & = & \frac{1}{s}
\end{eqnarray}
or from the equation $s \Theta_{0}(s) = r_1 \Theta_{1}(s)$, which is the Laplace transform of the equation (\ref{CME_WF}) for $n=0$.

We can invert (\ref{CME_WF_LT_solution_N}) and (\ref{CME_WF_LT_solution_n}) using the residue theorem. If the spectrum of the matrix $\mathbb{W}$ (\ref{CME_WF_matrix_form}) is non-degenerate, then all poles appearing in (\ref{CME_WF_LT_solution_n}) are single, i.e., of order 1. In such a case, we immediately obtain the equations (\ref{CME_WF_solution_N_t_main}) and (\ref{CME_WF_solution_n_t_main}).

%(Equivalently, the polynomial $\mathcal{M}_n(s) \equiv \prod_{j=n}^N (s + r_{j})$ has only single zeros.) 

\section{The derivation of the expressions for the first and second moments of \texorpdfstring{$P_n(t)$}{Pn(t)} \label{Derivation_of_moments_app}}

To derive Eqs. (\ref{first_moment_non_deg_WF}) and (\ref{second_moment_non_deg_WF}) we need several identities. The first one is an umbral variant of the binomial theorem, also known as the Chu-Vandermonde identity \cite{weisstein2010chuart}
\begin{equation}
\label{Chu_Vandermonde_identity}
(a+b)_m  = \sum_{j=0}^{m} {\binom{m}{j}} (a)_{m - j}(b)_j, 
\end{equation}
where $(u)_m$  denotes the Pochhammer symbol (the rising factorial),
\begin{equation}
\label{Pochhammer_symbol_definition}
(u)_m  = u(u+1)\cdots(u+m-1) = \frac{\Gamma(u+m)}{\Gamma(u)},
\end{equation}
and where $\Gamma(z)$ is the Gamma function \cite{abramowitz1968handbook}. The Pochhammer symbol has the property $(-x)_n =  (-1)^{n} (x-n+1)_n$. We also need the following result:
\begin{equation}
\label{Gamma_n_identity}
{\Gamma(z-n)} = (-1)^{n-1} \frac{\Gamma(-z)\Gamma(1+z)}{\Gamma(n+1-z)},  ~~~ n \in \mathbb{Z},~~~ z \notin \mathbb{Z},
\end{equation}
which can be derived from well-known identity \cite{abramowitz1968handbook}
%
%= \pi  \text{cosec}( \pi z)
%
\begin{equation}
{\Gamma(z)\Gamma(1-z)} = \frac{ \pi}{\sin( \pi z)},~~~ z \notin \mathbb{Z}.
\end{equation}
Our task now is to simplify the expression %for the two lowest moments, i.e., for
\begin{equation}
\label{q_th_moment_non_deg_WF_app}
   \mu_{qA}(t) \equiv \sum_{n=1}^N n^q P_{n}(t) = \sum_{n=1}^N n^q \sum_{k=n}^{N} C_{nk} e^{- r_{k} t} = \sum_{k=1}^{N} e^{- r_{k} t} \sum_{n=1}^k n^q  C_{nk}  
\end{equation}
for $q=1$ and $q=2$. To do this, in (\ref{q_th_moment_non_deg_WF_app}) we have to calculate the sum with respect to $n$, taking into account the explicit form (\ref{explicit_form_of_the_C_n_k_coefficients}) of the coefficients $C_{nk}$. For $q=1$ we get 
\begin{equation}
\label{n_sum_1_st_moment_non_deg_WF_app}
\sum_{n=1}^k n  C_{nk}  =   \frac{(-1)^{N+k}(2k - \tilde{M}) \Gamma(N+1)}{\Gamma(\tilde{M}-N)\Gamma(N - k + 1) \Gamma(N + k + 1 - \tilde{M})} \sum_{n=1}^k n \frac{\Gamma(\tilde{M}- n) \Gamma(n + k - \tilde{M})}{ \Gamma(k-n +1) \Gamma(n+1)}.
\end{equation}
In (\ref{n_sum_1_st_moment_non_deg_WF_app}) it is again the summation of the $n$-dependent part that we are interested in,
\begin{equation}
\label{n_sum_1_st_moment_non_deg_WF_app_sum_n_only}
\sum_{n=1}^k n \frac{\Gamma(\tilde{M}- n) \Gamma(n + k - \tilde{M})}{ \Gamma(k-n +1) \Gamma(n+1)} = \sum_{n=1}^k \frac{\Gamma(\tilde{M}- n) \Gamma(n + k - \tilde{M})}{ (k-n)!(n-1)!}.
\end{equation}
Now, making use of the definition (\ref{Pochhammer_symbol_definition}) of the Pochhammer symbol, we rewrite the numerator of (\ref{n_sum_1_st_moment_non_deg_WF_app_sum_n_only}) as
\begin{equation}
\label{n_sum_1_st_moment_non_deg_WF_app_sum_n_only_numerator}
\Gamma(\tilde{M}- n) \Gamma(n + k - \tilde{M}) = \Gamma(\tilde{M}- k) (\tilde{M}- k)_{k-1-(n-1)} \Gamma(k - \tilde{M} + 1) (k - \tilde{M} + 1)_{n-1}.
\end{equation}
Using (\ref{Gamma_n_identity}) we can replace $\Gamma(\tilde{M}- k)\Gamma(k - \tilde{M} + 1)$ by $(-1)^{k-1}\Gamma(-\tilde{M})\Gamma(\tilde{M} + 1)$. Then, by multiplying and dividing (\ref{n_sum_1_st_moment_non_deg_WF_app_sum_n_only}) by $(k-1)!$, we get
\begin{eqnarray}
\label{n_sum_1_st_moment_non_deg_WF_app_sum_n_only_bis}
\sum_{n=1}^k \frac{\Gamma(\tilde{M}- n) \Gamma(n + k - \tilde{M})}{ (k-n)!(n-1)!} &=& (-1)^{k-1} \frac{\Gamma(-\tilde{M})\Gamma(\tilde{M} + 1)}{(k-1)!} \sum_{n=1}^k  {\binom{k-1}{n-1}}  (\tilde{M}- k)_{k-1-(n-1)} (k - \tilde{M} + 1)_{n-1}.
% \nonumber \\ &=&.
\end{eqnarray}
By invoking (\ref{Chu_Vandermonde_identity}) we see that the sum over $n$ on the r.h.s. of Eq. (\ref{n_sum_1_st_moment_non_deg_WF_app_sum_n_only_bis}) is equal to $(1)_{k-1}=(k-1)!$, therefore 
\begin{eqnarray}
\label{n_sum_1_st_moment_non_deg_WF_app_sum_n_only_tris}
\sum_{n=1}^k \frac{\Gamma(\tilde{M}- n) \Gamma(n + k - \tilde{M})}{ (k-n)!(n-1)!} &=& (-1)^{k-1} \Gamma(-\tilde{M})\Gamma(\tilde{M} + 1).
% \nonumber \\ &=&.
\end{eqnarray}
Now again using (\ref{Gamma_n_identity}), we obtain (\ref{first_moment_non_deg_WF}) from (\ref{n_sum_1_st_moment_non_deg_WF_app}). The derivation of (\ref{second_moment_non_deg_WF}) is similar. It is convenient to compute  $\sum_{n=2}^N n(n-1) P_{n}(t)$ (i.e., the second factorial moment) instead of $\mu_{2A}(t) = \sum_{n=1}^N n^2 P_{n}(t)$ and use the expression for $\mu_{1A}(t)$ to finally get $\mu_{2A}(t)$.

\section{Rate equations of the deterministic kinetics and their solutions}

\subsection{The original Finke--Watzky model (\ref{1st_reaction_WF})--(\ref{2nd_reaction_WF}) \label{relation_of_sto_to_det}}

For reactions (\ref{1st_reaction_WF})–(\ref{2nd_reaction_WF}), the deterministic kinetic rate equations have the following form:
\begin{eqnarray}
\label{Det_Chem_Kin_WF_system_A_main}
\frac{d a}{d t} &=& - \mathcal{K}_{1} a - \mathcal{K}_{2} a b, \\
\label{Det_Chem_Kin_WF_system_B_main}
\frac{d b}{d t} &=&   \mathcal{K}_{1} a + \mathcal{K}_{2} a b,
\end{eqnarray}
where $a(t)$ and $b(t)$ denote the concentrations of A and B, respectively, and the initial condition is $a(0) = a_0$, $b(0) = b_0$. From (\ref{Det_Chem_Kin_WF_system_A_main})–(\ref{Det_Chem_Kin_WF_system_B_main}) it follows that $a(t) + b(t) = a_0 + b_0$ (this condition corresponds to Eq.~(\ref{constraint_on_mass_conservation})), so these two equations can be replaced by a single one:
\begin{eqnarray}
\label{Det_Chem_Kin_WF_only_a}
\frac{d a}{d t} &=& - a \left[ \mathcal{K}_{1} + \mathcal{K}_{2} (a_0 + b_0) - \mathcal{K}_{2} a\right]. 
\end{eqnarray}
The solution of (\ref{Det_Chem_Kin_WF_only_a}),
\begin{eqnarray}
\label{Det_Chem_Kin_WF_only_a_solution_of}
a(t) &=& a_0 \frac{\mathcal{K}_{1} + \mathcal{K}_{2}(a_0 + b_0) }{\mathcal{K}_{2}a_0  + (\mathcal{K}_{1} + \mathcal{K}_{2}b_0)  \exp \{ [\mathcal{K}_{1} + \mathcal{K}_{2}(a_0 + b_0)]t\} },  \nonumber \\
\end{eqnarray}
is used to fit experimental data; see, e.g., Refs.~\onlinecite{watzky1997transition, morris2008fitting, morris2009protein, iashchishyn2017finke}. The stationary solution of (\ref{Det_Chem_Kin_WF_only_a_solution_of}) is
\begin{equation}
\lim_{t \to \infty} a(t) = 0, \quad \lim_{t \to \infty} b(t) = a_0 + b_0,
\end{equation}
which agrees with its stochastic counterpart (\ref{deterministic_final_condition}). The rate constants $\mathcal{K}_{1}$ and $\mathcal{K}_{2}$ in Eqs.~(\ref{Det_Chem_Kin_WF_system_A_main})–(\ref{Det_Chem_Kin_WF_only_a_solution_of}) correspond to the CME rate constants $k_1$ and $k_2$ in Eq.~(\ref{CME_WF}). While $\mathcal{K}_{1} = k_1$, the constant $\mathcal{K}_{2}$ generally differs from $k_2$ (see Eq.~\ref{relationship_K_k_rate_constants}). Note also that $\mathcal{K}_{1}$ and $\mathcal{K}_{2}$ have different dimensions.

\subsection{Reversible generalization of the Finke-Watkzy model (\ref{1st_reaction_WF_rev})--(\ref{2nd_reaction_WF_rev}) \label{Rate_equations_rev_App}}

For the set of chemical reactions (\ref{1st_reaction_WF_rev})-(\ref{2nd_reaction_WF_rev}) the determnisitic kinetic rate equations read
\begin{eqnarray}
\label{Det_Chem_Kin_WF_system_A_rev}
\frac{d a}{d t} &=& - \mathcal{K}^{+}_{1} a - \mathcal{K}^{+}_{2} a b + \mathcal{K}_{1}^{-} b  + \mathcal{K}^{-}_{2} b^2, \\
\label{Det_Chem_Kin_WF_system_B_rev}
\frac{d b}{d t} &=&   \mathcal{K}_{1}^{+} a + \mathcal{K}^{+}_{2} a b -  \mathcal{K}_{1}^{-} b - \mathcal{K}^{-}_{2} b^2.
\end{eqnarray}
"Deterministic" kinetic reaction rate constants $\mathcal{K}^{\pm}_{1}$, $\mathcal{K}^{\pm}_{2}$ appearing in the above equations correspond to the "stochastic" rate constants $k^{\pm}_1$ and $k^{\pm}_2$ of the CME (\ref{CME_WF_rev}) with $r_{n}$ (\ref{r_n_of_CME_WF_rev}) and $g_{n}$ (\ref{g_n_of_CME_WF_rev}), see Eq. (\ref{relationship_K_k_rate_constants_rev}). As in the case of Eq. (\ref{Det_Chem_Kin_WF_system_A_main})-(\ref{Det_Chem_Kin_WF_system_B_main}), here we again have $a(t) + b(t) = a(0) + b(0) \equiv \mathcal{M}$. Therefore instead of (\ref{Det_Chem_Kin_WF_system_A_rev})-(\ref{Det_Chem_Kin_WF_system_B_rev}) we can use a single equation
\begin{eqnarray}
\label{Det_Chem_Kin_WF_only_a_rev}
\frac{d a}{d t} &=& - \mathcal{K}^{+}_{1} a - \mathcal{K}^{+}_{2} a (\mathcal{M} - a) + \mathcal{K}_{1}^{-} (\mathcal{M} - a)  + \mathcal{K}^{-}_{2} (\mathcal{M} - a)^2 \nonumber \\ &=& ~~(\mathcal{K}^{+}_{2} + \mathcal{K}^{-}_{2})(a-a_{+})(a-a_{-}),
\end{eqnarray}
%
%\nonumber \\  &=&
%
where $a_{\pm}$ are given by
\begin{eqnarray}
\label{Det_Chem_Kin_WF_only_a_rev_roots}
a_{\pm} &=& \frac{\mathcal{K}^{+}_{1} + \mathcal{K}^{-}_{1} + \mathcal{M}(\mathcal{K}^{+}_{2} + 2 \mathcal{K}^{-}_{2}) \pm \sqrt{[\mathcal{K}^{+}_{1} + \mathcal{K}^{-}_{1} + \mathcal{M}(\mathcal{K}^{+}_{2} + 2 \mathcal{K}^{-}_{2})]^2 - 4(\mathcal{K}^{+}_{2} + \mathcal{K}^{-}_{2})(\mathcal{K}^{-}_{1} + \mathcal{M}\mathcal{K}^{-}_{2})\mathcal{M} }}{2(\mathcal{K}^{+}_{2} + \mathcal{K}^{-}_{2})}.
\end{eqnarray}
We see that there are two stationary solutions of  (\ref{Det_Chem_Kin_WF_only_a_rev}) as given by   (\ref{Det_Chem_Kin_WF_only_a_rev_roots}), but  only $a_{-}$ is the physical one, as $a_{+}> \mathcal{M}$. It is clear that for $\mathcal{K}^{-}_{1} = \mathcal{K}^{-}_{2} = 0$ ($k^{-}_{1} = k^{-}_{2} = 0$), i.e. if the reaction network (\ref{1st_reaction_WF_rev})-(\ref{2nd_reaction_WF_rev}) reduces to FWM (\ref{1st_reaction_WF})-(\ref{2nd_reaction_WF}), we have $a_{-}=0$.

The time-dependent solution of (\ref{Det_Chem_Kin_WF_only_a_rev}) can be written in the following form:
\begin{eqnarray}
\label{Det_Chem_Kin_WF_only_a_solution_of_rev}
a(t) &=& 
\frac{
  a_{+} - a_{-} \, \displaystyle\frac{a_{0}-a_{+}}{a_{0}-a_{-}}
  \, e^{\,(\mathcal{K}_{2}^{+} + \mathcal{K}_{2}^{-})\,(a_{+}-a_{-})\,t}
}{
  1 - \displaystyle\frac{a_{0}-a_{+}}{a_{0}-a_{-}}
  \, e^{\,(\mathcal{K}_{2}^{+} + \mathcal{K}_{2}^{-})\,(a_{+}-a_{-})\,t}
},
\end{eqnarray}
where $a_{0} = a(0)$ is the initial concentration and 
$a_{\pm}$ are given by Eq. (\ref{Det_Chem_Kin_WF_only_a_rev_roots}).

\subsection{The inverse Finke–Watzky model (\ref{1st_reaction_inv_WF})–(\ref{2nd_reaction_inv_WF})}

The deterministic rate equation for \(a(t)\) reads
\begin{equation}
\label{Det_Chem_Kin_inv_WF_only_a}
\frac{da}{dt}=-a\bigl(\mathcal{K}_1+\mathcal{K}_2\,a\bigr),
\end{equation}
with the solution
\begin{equation}
\label{Det_Chem_Kin_inv_WF_only_a_solution_of}
a(t)
=\frac{a_0}
     {\bigl(\tfrac{\mathcal{K}_2}{\mathcal{K}_1}a_0+1\bigr)
      e^{\mathcal{K}_1t}-\tfrac{\mathcal{K}_2}{\mathcal{K}_1}a_0}.
\end{equation}

The relationship between deterministic and stochastic rate constants differs from that in the original FWM. In the present case we have
\begin{equation}
\label{relationship_K_k_rate_constants_inv}
k_1 = \mathcal{K}_{1}, \qquad k_2 = \frac{2 \mathcal{K}_{2}}{V}.
\end{equation}

\end{widetext}

%\bibliography{bibliographyCMEs_2}  

%%%%%%%%%%%%%%

\bibliography{bibliography_CME_WF_theor}   % Produces the bibliography via BibTeX.

\end{document}